\documentclass[manuscript]{aastex}

\usepackage{url}
\shorttitle{IRIS observations of a solar flare}
\shortauthors{Sadykov et al.}

\begin{document}

\title{PROPERTIES OF CHROMOSPHERIC EVAPORATION AND PLASMA DYNAMICS OF A SOLAR FLARE FROM IRIS OBSERVATIONS}

\author{Viacheslav M. Sadykov\altaffilmark{1,2,3,4}, Santiago Vargas Dominguez\altaffilmark{1,5}, Alexander G. Kosovichev\altaffilmark{1,2}, Ivan N. Sharykin\altaffilmark{3}, Alexei B. Struminsky\altaffilmark{3,4} and Ivan Zimovets\altaffilmark{3}}
\affil{$^1$Big Bear Solar Observatory, New Jersey Institute of Technology, Big Bear City, CA 92314, USA}
\affil{$^2$NASA Ames Research Center, Moffett Field, CA 94035, USA}
\affil{$^3$Space Research Institute (IKI) of Russian Academy of Sciences, Moscow 117997, Russia}
\affil{$^4$Moscow Institute of Physics and Technology (MIPT), Dolgoprudny, Moscow Region 141700, Russia}
\affil{$^5$Observatorio Astron\'omico Nacional Universidad Nacional de Colombia, Bogot\'a, Colombia}

\begin{abstract}
Dynamics of hot chromospheric plasma of solar flares is a key to understanding of mechanisms of flare energy release and particle acceleration. A moderate M1.0 class flare of 12 June, 2014 (SOL2014-06-12T21:12) was simultaneously observed by NASA's Interface Region Imaging Spectrograph (IRIS), other spacecraft, and also by New Solar Telescope (NST) at the BBSO. This paper presents the first part of our investigation focused on analysis of the IRIS data. Our analysis of the IRIS data in different spectral lines reveals strong redshifted jet-like flow with the speed of $\sim$100\,km/s of the chromospheric material before the flare. Strong nonthermal emission of the C\,II\,k\,1334.5\,{\AA} line, formed in the chromosphere-corona transition region, is observed at the beginning of the impulsive phase in several small (with a size of $\sim$1\,arcsec) points. It is also found that the C\,II\,k line is redshifted across the flaring region before, during and after the impulsive phase. A peak of integrated emission of the hot ($1.1\cdot 10^{7}$\,K) plasma in the Fe\,XXI\,1354.1\,{\AA} line is detected approximately 5 minutes after the integrated emission peak of the lower temperature C\,II\,k. A strong blueshift of the Fe\,XXI line across the flaring region corresponds to evaporation flows of the hot chromospheric plasma with a speed of 50\,km/s. Additional analysis of the Reuven Ramaty High-Energy Solar Spectroscopic Imager (RHESSI) data supports the idea that the upper chromospheric dynamics observed by IRIS has features of ``gentle'' evaporation driven by heating of the solar chromosphere by accelerated electrons and by a heat flux from the flare energy release site.
\end{abstract}


\section{Introduction}

Spectroscopic observations of solar flares are very important and necessary for understanding the flare energetics and dynamics. Different spectral lines (and even different parts of a spectral line, e.g. core and wings) have different effective formation heights and allow us to obtain the information about the atmospheric structure and parameters: temperature, line-of-sight velocities, number density, etc. \citep{Leenaarts12,Leenaarts13a}. Thus, spectroscopic observations help us to study different plasma flows. For example, \citet{Tian14b} studied the properties of plasma jets from a reconnection region. Another phenomenon connected with plasma flows is a chromospheric evaporation in solar flares. There are many observations of this phenomenon \citep{Brosius04,Milligan06,Doschek13}. A recent overview of observations of solar flares in general is published by \citet{Fletcher11}, and the chromospheric evaporation processes are reviewed by \citet{Milligan15}. However, the basic mechanisms of the plasma evaporation and the heating sources causing this evaporation are not yet understood.

Recently it became possible to study plasma flows in the solar chromosphere and transition region with the NASA Interface Region Imaging Spectrograph (IRIS) mission \citep{DePontieu14}. The IRIS mission allows us to obtain and analyze spectral line profiles of the chromospheric and transition region plasma. The instrument obtains UV spectra in three wavelength diapasons, and also slit-jaw images with four filters with high spectral (26-53\,m{\AA}), spatial (0.33 arcsec), and temporal ($\approx$2\,sec) resolutions. This allows us to look in detail at the spectral characteristics of each flare point, and provides us with an opportunity to understand the fine structure of individual energy release event.

In this work, we present initial results of analysis of IRIS observations of the 12 June, 2014, M1.0 class flare event, the soft X-ray emission peak of which occurred at about 21:12\,UT. This event was simultaneously observed by IRIS and the 1.6\,m New Solar Telescope (NST). Here we present details of our analysis of the IRIS data complemented by X-ray data from GOES and RHESSI \citep{Lin02}. The X-ray spectral data provide information about accelerated particles and coronal heating. Analysis of the NST observations will be presented in a subsequent paper. We also use data from the Solar Dynamic Observatory (SDO) Helioseismic and Magnetic Imager (HMI) \citep{Scherrer12} and Atmospheric Imaging Assembly (AIA) \citep{Lemen12}.

\section{General Description of Observations and Methods}

\begin{figure}[t]
\centering
\includegraphics[width=1.0\linewidth]{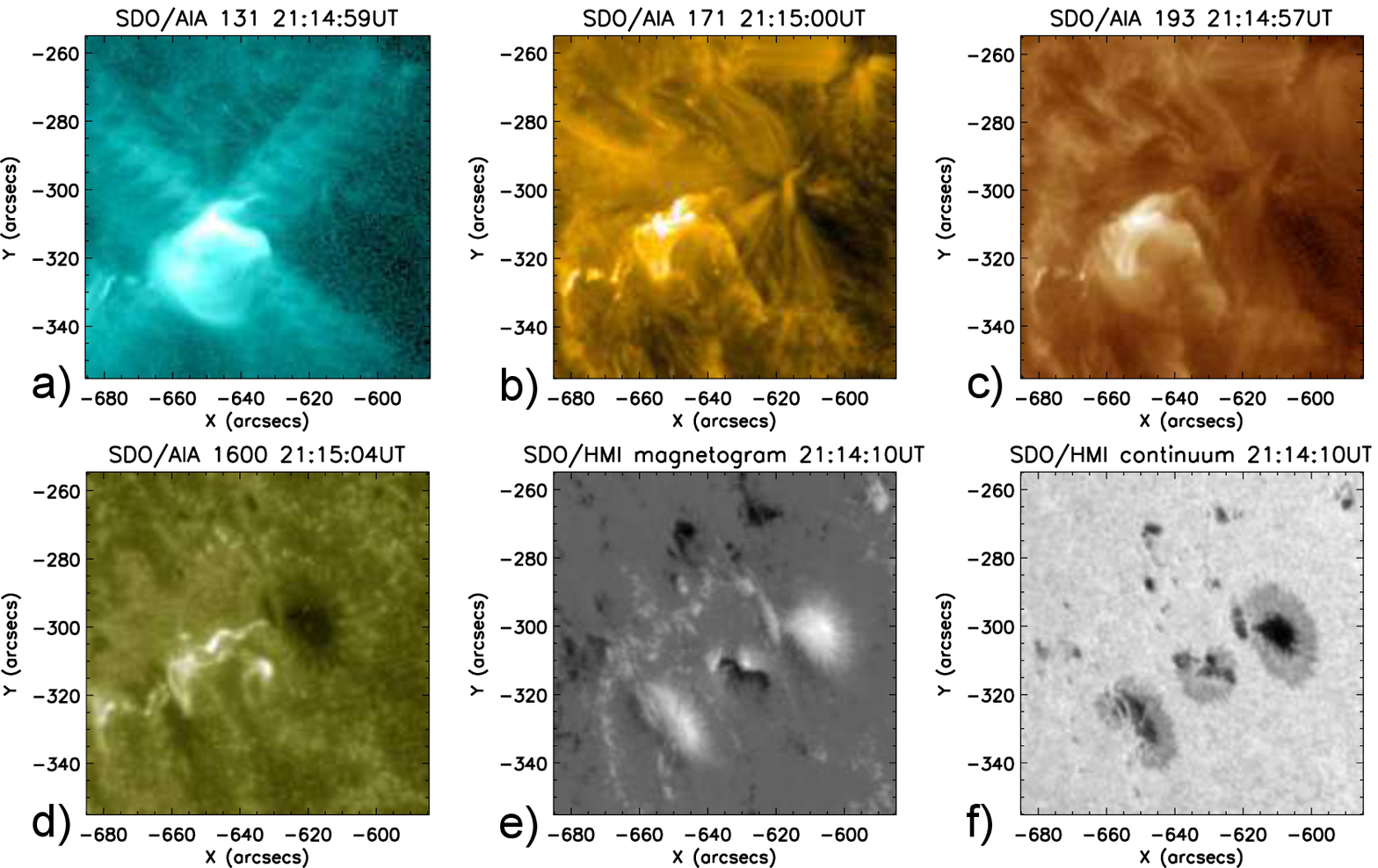}
\caption{Context SDO images for the M1.0 class flare event of 12 June, 2014, obtained in the AIA EUV channels a)~131\,{\AA}, b)~171\,{\AA}, c)~193\,{\AA} and d)~1600\,{\AA}, and e)~the HMI magnetogram and f)~continuum intensity image. The flare was observed in the central $\delta$-type sunspot.}
\label{figure1}
\end{figure}

\begin{figure}[t]
\centering
\includegraphics[width=0.65\linewidth]{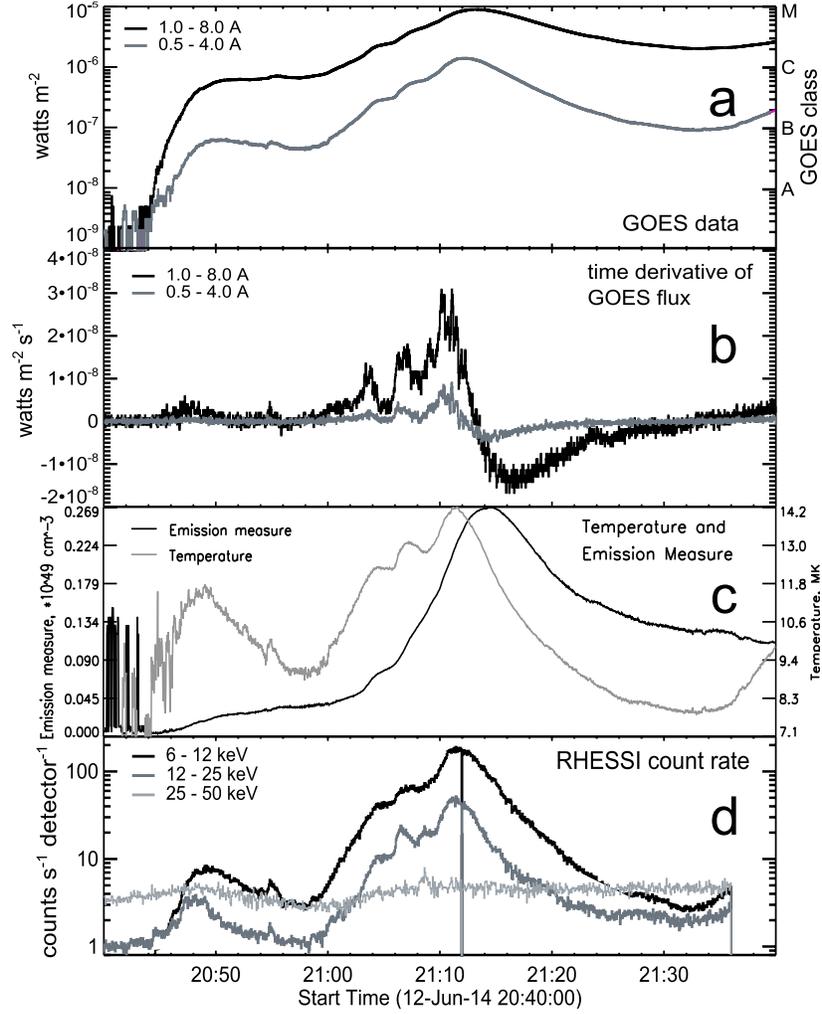}
\caption{Properties of the flare X-ray emission: a)~soft X-ray light curves from the GOES satellite, b)~time derivatives of the soft X-ray flux, c)~the temperature and emission measure estimated from the GOES data, d)~lightcurves of the X-ray emission in three energy bands: 6-12, 12-25 and 25-50\,keV from the RHESSI satellite.}
\label{figure2}
\end{figure}

The M1.0 class flare occurred on 12 June, 2014, at around 21:12\,UT in active region NOAA 12087, located approximately 50 degrees South-East from the disk center. In Figure~\ref{figure1}, a series of context SDO images of this region are displayed. This region was very active and produced several M-class flares and one X-class flare during the period of 11-13 June, 2014. The X-ray fluxes observed by the GOES X-ray sensor (XRS) \citep{Bornmann96}, and the RHESSI \citep{Lin02} satellite are shown in Figure~\ref{figure2}. The flare peak near 21:12UT is observed in both GOES channels, and in the RHESSI 6-12\,keV and 12-25\,keV channels. The hard X-ray flux ($>$~25\,keV) was very weak during the event. The X-ray fluxes in the energy ranges 6-12\,keV and 12-25\,keV are sufficiently strong for applying a reconstruction procedure to determine the X-ray emission centroids. We also study the RHESSI spectra in the energy range 8-25 keV, using 1,3,6,8, and 9 RHESSI detectors, and applying a least squares fitting procedure from the OSPEX Solar SoftWare (SSW) package. The standard SSW procedures are applied to calculate the temperature and emission measure based on the soft X-ray data from the GOES satellite (hereafter, we refer to these estimates as ``GOES temperature'' and ``GOES emission measure''). The algorithm used in these procedures is described by \citet{Thomas85}. The GOES temperature and emission measure curves are also displayed in Figure~\ref{figure2}.

IRIS obtained good spatial and temporal coverage of the flare event. The instrument observed the central $\delta$-type spot of the NOAA 12087 active region from 18:44:38 UT to 23:59:44 UT on 12 June, 2014 in the coarse raster mode (8 slit positions covering the region of interest). The Slit-Jaw (SJ) images with two filters (Mg\,II\,k 2796 and C\,II 1330), and the spectral data in 9 different intervals, including the Mg\,II\,h\&k, Si\,IV, C\,II and O I lines, are available for this event. We consider the dynamics of the region from 20:15:00\,UT until 22:00:00\,UT. Figure~\ref{figure3} displays the flare evolution in the IRIS SJ images. The slit positions are indicated by white lines.

The optically thick Mg\,II\,h\&k spectral lines usually represent two-peak structures in quiet Sun regions. Correlations between characteristics of these lines and physical parameters of the atmosphere are discussed in the paper of \citet{Leenaarts13a}. The authors used radiative-MHD BIFROST simulation results \citep{Gudiksen11} and computed synthetic spectra. However, the appearance of the Mg\,II\,h\&k lines in the flare observations is completely different: as we will see further, the lines have Gaussian-type profiles without any significant two-peak structures. It is necessary to mention that the absence of self-reversals (two-peak structure) of the Mg\,II lines was also observed in sunspots \citep[see Figure~4 of][]{Tian14a} and in prominences \citep[see Figure~8 of][]{Schmieder14}.

For analysis of the Mg\,II, and also C\,II and Si\,IV lines, we consider three parameters: 1) maximum of the line intensity, 2) the Doppler shift estimate defined as a difference between the center of gravity of the line and the reference wavelength for this line $ \left<{\lambda}\right> - \lambda{}_{ref}= {\int{{\lambda}Id{\lambda}}}/{\int{Id{\lambda}}} - \lambda{}_{ref} $, 3) the line halfwidth estimated as $\sqrt{\left<(\lambda-\bar{\lambda})^{2}\right>} = \sqrt{{\int{{\lambda}^{2}Id{\lambda}}}/{\int{Id{\lambda}}}-\bar{\lambda}^{2}} $. We also study in detail behavior of the line profiles in several characteristic flare points.

The flare dynamics is very complex. During an initial time interval, from 20:15:00\,UT until 20:48:00\,UT, the IRIS images show a prominent loop-like structure located at the (-650$^{\prime\prime}$,-310$^{\prime\prime}$) point near the left-most slit position. This structure is not spatially covered by the IRIS raster positions, and therefore there is no information about its spectra. The second time interval that can be distinguished in the pre-flare development lasts from 20:48:00\,UT until approximately 21:05:00\,UT. During this period the brightest structure is a strongly redshifted jet that emerges from the loop-like region mentioned above. At $\sim$21:05:00\,UT the impulsive flare phase begins. A sharp increase in brightness of a large area is evident in both SJ channels. The maximum brightness for most of the flare region occurs between 21:10:00\,UT and 21:15:00\,UT. After this, the decaying phase of the flare begins.

\section{Detailed Analysis of Observations}

\subsection{Strongly Redshifted Plasma Jet}

\begin{figure}[t]
\centering
\includegraphics[width=0.83\linewidth]{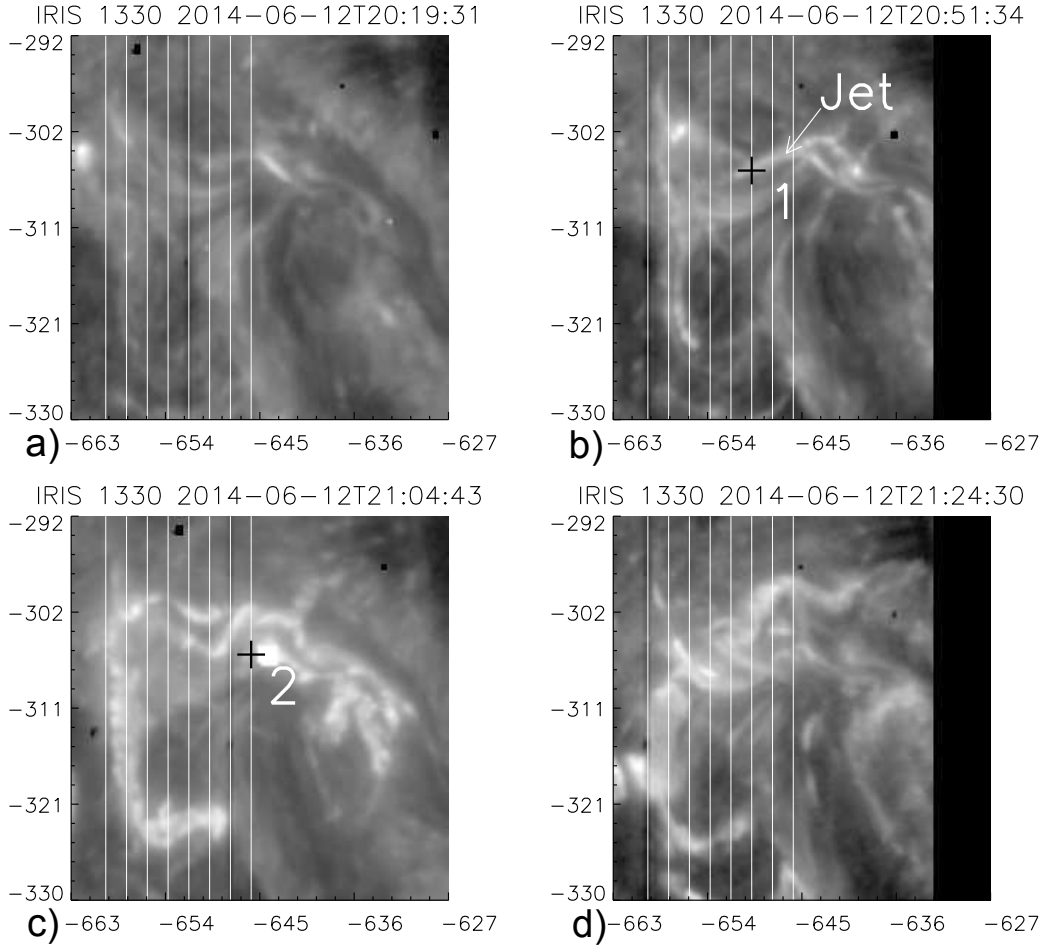}
\caption{Evolution of the 1330\,{\AA} intensity obtained from IRIS slit-jaw imager for four moments of time before, during and after the flare: a)~20:19:31\,UT, b)~20:51:34\,UT, c)~21:04:43\,UT, d)~21:24:30\,UT. The IRIS slit positions are shown by the white vertical lines. Black crosses mark the location of: 1)~the characteristic jet point; 2)~the characteristic flare point. The image coordinates are the disk coordinates relative to the solar disc center in arcseconds.}
\label{figure3}
\end{figure}

The strongly redshifted jet is observed in the region just prior the flare. The jet appears as a very bright, approximately 6\,Mm long structure in the IRIS SJ images. It starts from the bright loop-like structure (see IRIS images for 20:51:34\,UT in Figure~\ref{figure3}) and exists until the beginning of the flare. Approximate coordinates of this jet are (-650$^{\prime\prime}$,-310$^{\prime\prime}$) in the disk coordinate system. The jet structure is non-uniform and consists of numerous tiny brightenings traveling South-East. Therefore, it is likely that the jet plasma also flows in this direction. The four leftmost IRIS slit positions cross the jet, and allow us to analyze its spectral characteristics. We choose one of the slit positions crossing the jet (marked by a black cross and number 1 in Figure~\ref{figure3}) for a detailed analysis.

\begin{figure}[t]
\centering
\includegraphics[width=0.6\linewidth]{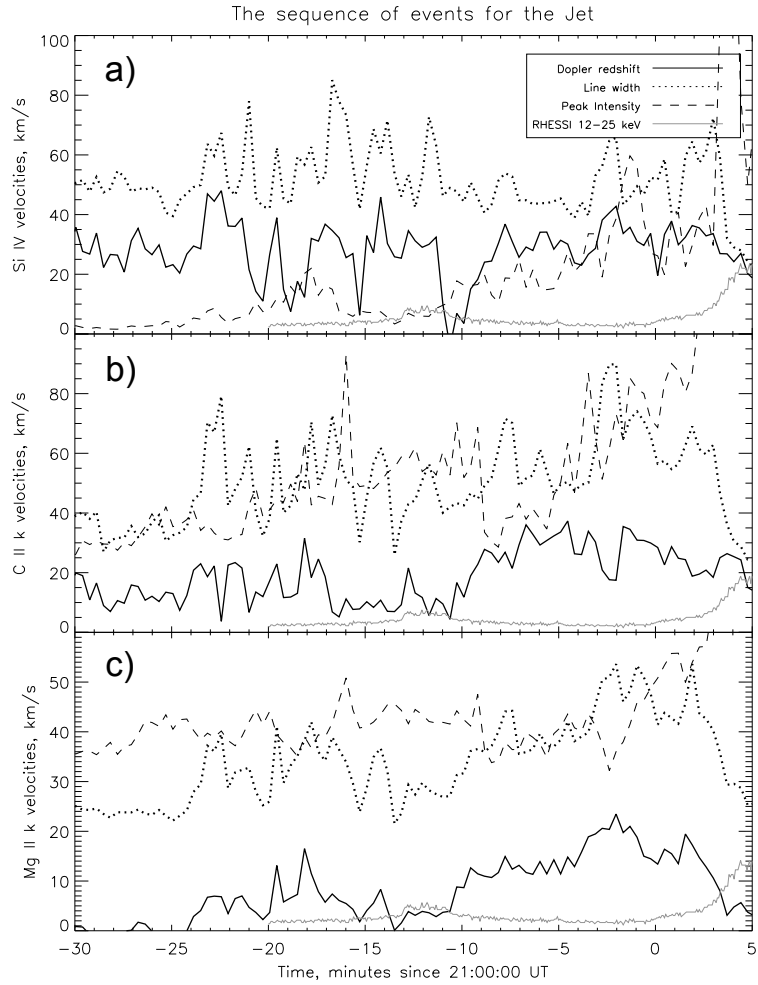}
\caption{Time evolution of the Doppler shift (black solid), line width (black dotted) and peak intensity (black dashed) for the jet event for: a)~Si\,IV, b)~C\,II\,k, and c)~Mg\,II\,k lines. The 12-25\,keV X-ray light curve (gray solid) is overplotted in all three panels.}
\label{figure4}
\end{figure}

Figure~\ref{figure4} presents the peak intensities, Doppler-shift and line-halfwidth estimates for three IRIS lines: Si\,IV (1402.82\,{\AA}), C\,II\,k (1334.55\,{\AA}), and Mg\,II\,k (2796.35\,{\AA}). The reference wavelengths of the lines are obtained from the atomic spectral line database \citep{Kurucz95}, and additionally calibrated more precisely at several quiet Sun points. The time scale in Figure~\ref{figure4} is shown relative to 21:00:00 UT and is measured in minutes. The Doppler shift estimate (positive for a redshift) is shown in km/s, and the line-width measure is presented in the same units. The peak line intensities are normalized to the maximum intensity. The X-ray flux in the RHESSI 12-25\,keV channel is also plotted for comparison. This figure shows that the jet is redshifted, on average, and that there are some variations of the Doppler shift at the jet point. These variations are also observed in the line-width estimates and in the peak intensity. This behavior means that the plasma in the jet is non-uniform and has a fine structure, as we mentioned before. The jet probably consists of series of injections initiated from the bright loop-like region. However, the average Doppler shift starts to increase at 20:46:35\,UT, and at this time the jet becomes more homogeneous, as also seen in the SJ images.

\begin{figure}
\centering
\includegraphics[width=0.9\linewidth]{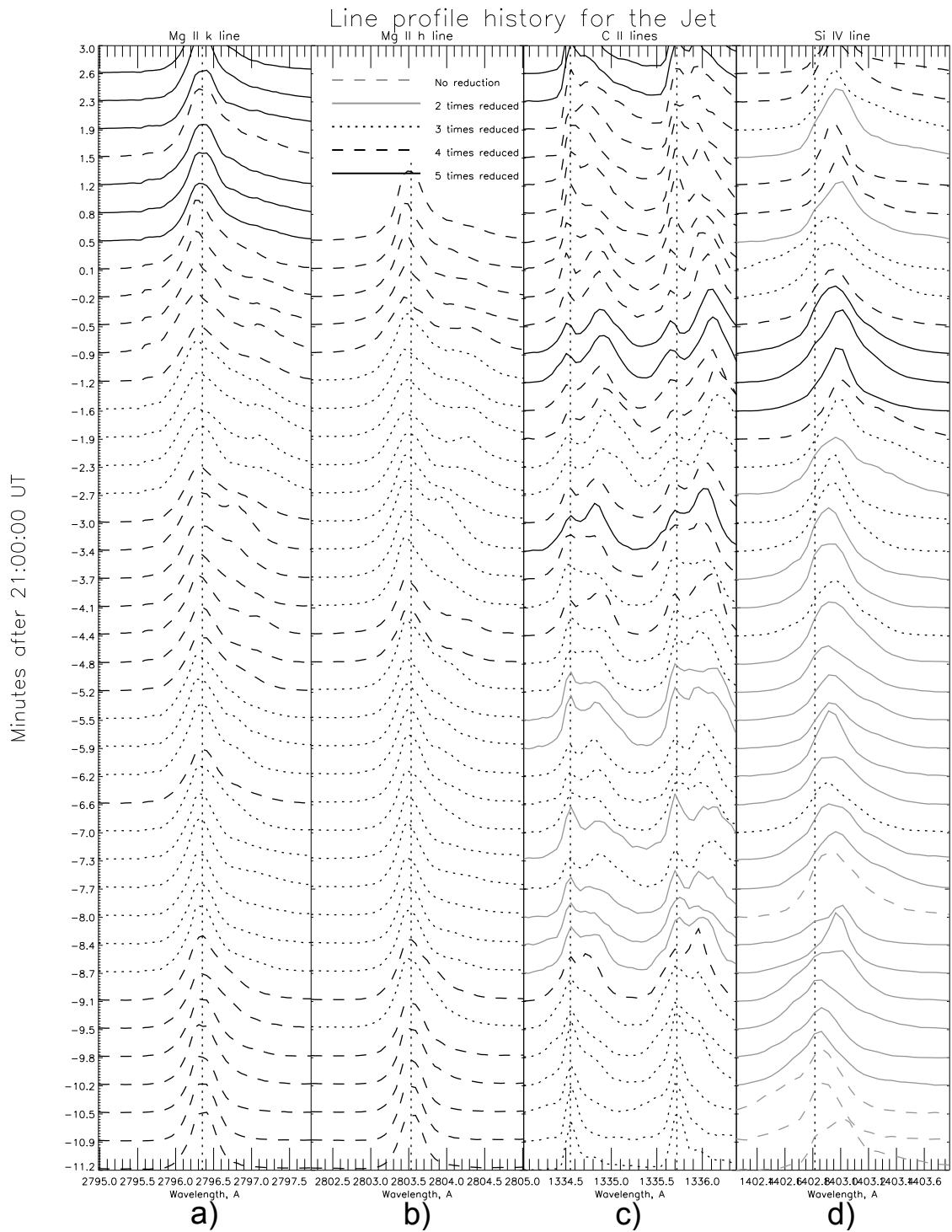}
\caption{Line profile evolution for the jet event. Profiles are presented for: a)~Mg\,II\,k, b)~Mg\,II\,h, c)~C\,II\,h\&k, and d)~Si\,IV lines. The gray scale and line style correspond to the scaling factors (see text and legends at the top of panel b) for details).}
\label{figure5}
\end{figure}

Figure~\ref{figure5} shows the line profiles evolution for the jet event. For convenience, the line profiles are plotted with different styles for different amplitudes, and their amplitude is scaled according to the following style code scheme: if the line peak intensity is greater than 80\% of the maximum peak intensity of the whole set, the profile is scaled by a factor of 5 and presented by a black solid line; from 60\% to 80\%, the line profile is a black dashed line and scaled by a factor of 4, and so forth (see notations in Figure~\ref{figure5} for details). The jet represents an enhancement of the red wing of the line, and sometimes appears as an additional peak (which is clearly observed, for example, in the profiles corresponding to 20:58\,UT). Such a line profile is likely due to a two-component plasma along the line of sight: one component belongs to the jet, and another one belongs to the background plasma. The typical Doppler velocities corresponding to the redshifted peak are approximately 100\,km/s, which is significantly higher than the Doppler shift estimate based on the center of gravity. This difference is easy to explain because the two plasma components have different red shifts, and the Doppler shift estimate represents superposition of these.

The origin of the preflare jet is unclear. Perhaps, it represents a downward plasma flow from the region of magnetic reconnection. Since the region is located $\sim$50 degrees from the disc center, the actual plasma velocities may be higher. The appearance of the second peak in the line profile at jet point indicates the presence of a two-component plasma along the line of sight. This is not the first spectroscopic observation of complex line profiles indicating a two-component plasma. For example, such results were previously reported in works of \citet{Chifor08} and \citet{Tian12}. Analysis of the NST observations, which we plan to present in our next paper, can shed light to the origin of this jet, and also can help us to understand its structure and evolution.

\subsection{Impulsive Phase of the Flare}

\begin{figure}[t]
\centering
\includegraphics[width=0.75\linewidth]{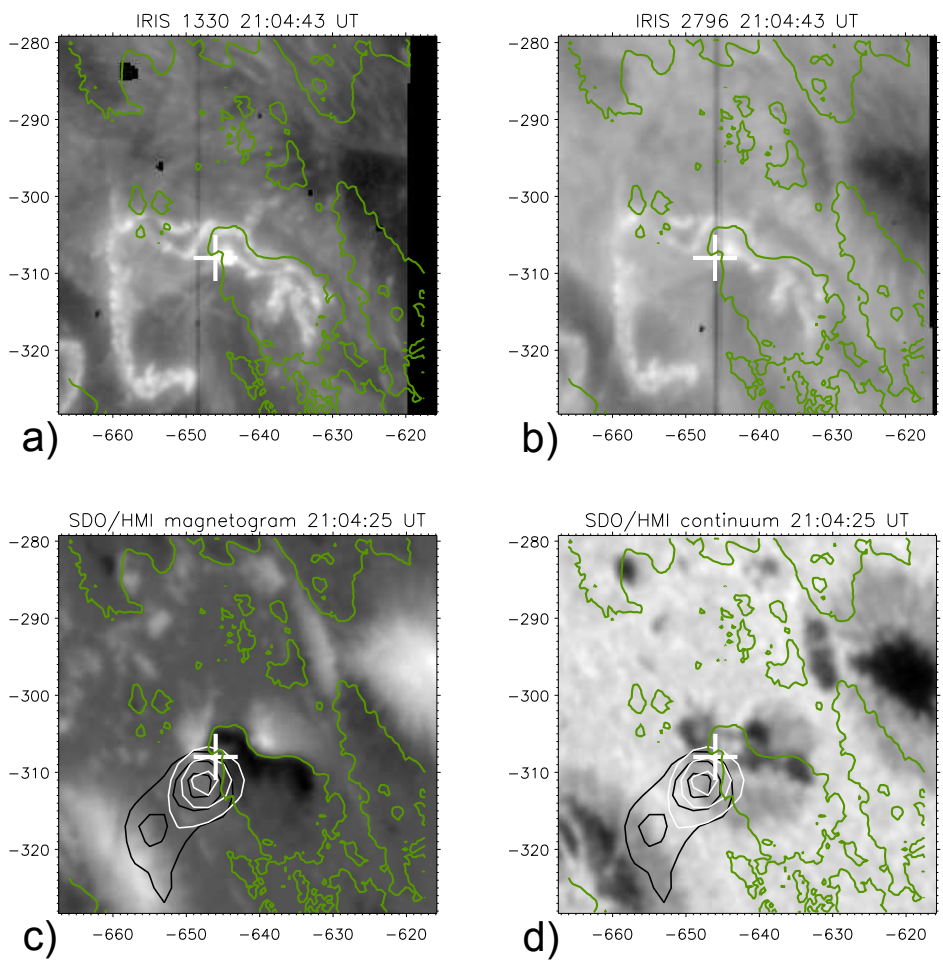}
\caption{Images of the flare region in 1330\,{\AA}~(a) and 2796\,{\AA}~(b) lines, aligned with the corresponding HMI line-of-sight magnetogram~(c) and continuum image~(d). Reconstructed from the RHESSI data the 12-25\,keV (black) and 6-12\,keV (white) X-ray sources for the time interval of 21:04:00-21:06:00\,UT, and the magnetic line-of-sight polarity inversion line (green) are overplotted. The X-ray contours correspond to 0.9, 0.7 and 0.5 of the maximum level. White cross indicates the characteristic flare point discussed in the text.}
\label{figure6}
\end{figure}

The M1.0 class flare appears in the IRIS SJ images as a strong and spatially broad brightening. Figure~\ref{figure6} shows images in the IRIS 1330\,{\AA} and 2796\,{\AA} lines, aligned with the simultaneous HMI line-of-sight magnetogram and continuum images. The reconstructed X-ray RHESSI flux centroids for the 6-12\,keV and 12-25\,keV energy ranges, integrated over the 21:04:00\,UT-21:06:00\,UT time interval, are shown by contour lines. The magnetogram shows that the central spot where the flare occurred has a very complex $\delta$-type magnetic field structure. The above-mentioned loop-like structure is located near the polarity inversion line.

\begin{figure}[t]
\centering
\includegraphics[width=0.7\linewidth]{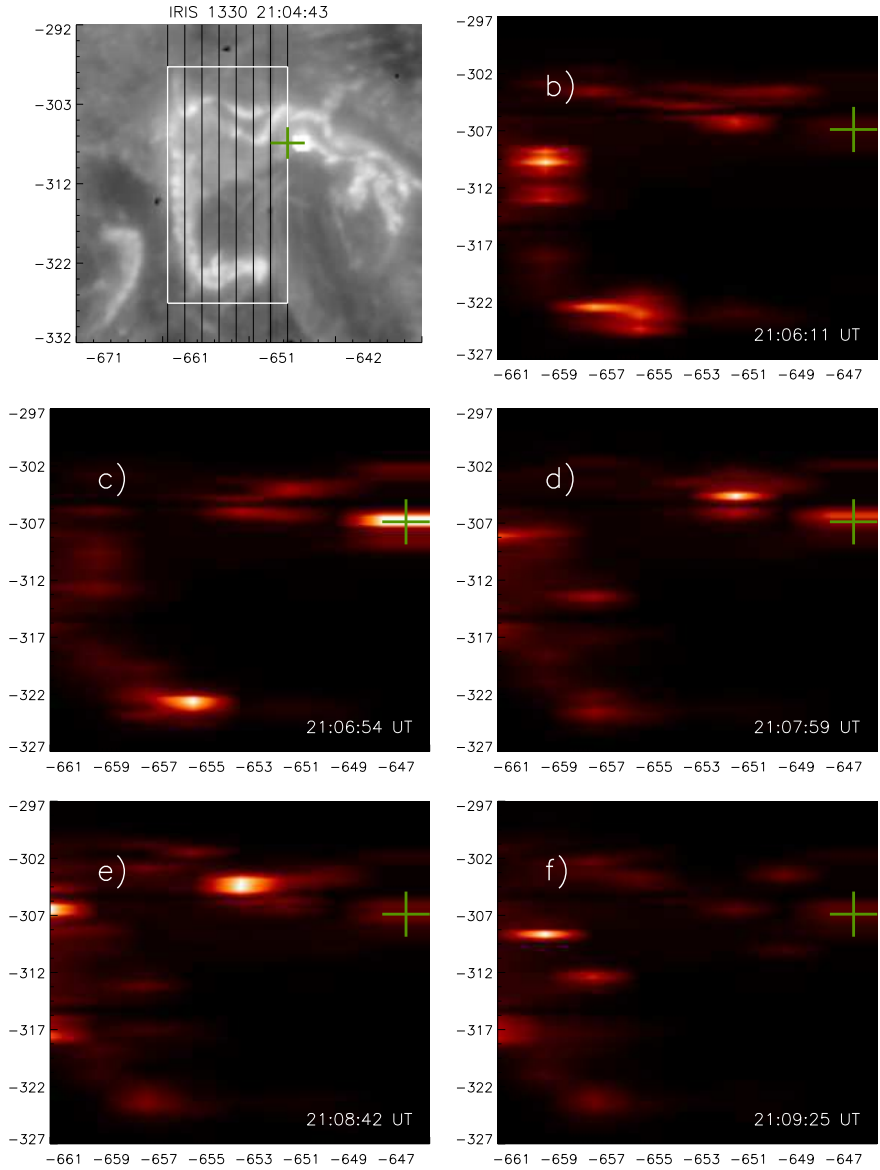}
\caption{Time evolution of the peak intensity map for the C\,II\,k line. Description of images: a) IRIS 1330\,{\AA} slit-jaw image at 21:04:43\,UT with the marked slit positions (black vertical lines) and the characteristic flare point (green cross). The white rectangle corresponds to the region, for which peak intensity maps are plotted in panels b)-f) for five moments of time marked in the lower right corner of each image.}
\label{figure7}
\end{figure}

Figure~\ref{figure7} displays a series of intensity maps for the C\,II\,k line peak corresponding to shorter wavelengths. The maps represent the peak intensity determined at 8 slit positions with the 0.33 arcsec resolution along the slits, and linearly interpolated between the slit positions. The selected time moments are marked in the bottom right corner of the images. These maps cover the time moments when the peak intensity of the C\,II\,k line had significant brightenings across the flaring region. Notice that these brightenings are localized, with the spatial size of about 1\,arcsec, and occur at various points prior to the X-ray peak of the flare. For a detailed analysis of the spectra we consider the point marked by green cross in Figure~\ref{figure7} (hereafter, the characteristic flare point). This point (marked with number 2 in Figures~\ref{figure3}~and~\ref{figure6}) is located near the 12-25\,keV X-ray emission centroid.

\begin{figure}[t]
\centering
\includegraphics[width=0.59\linewidth]{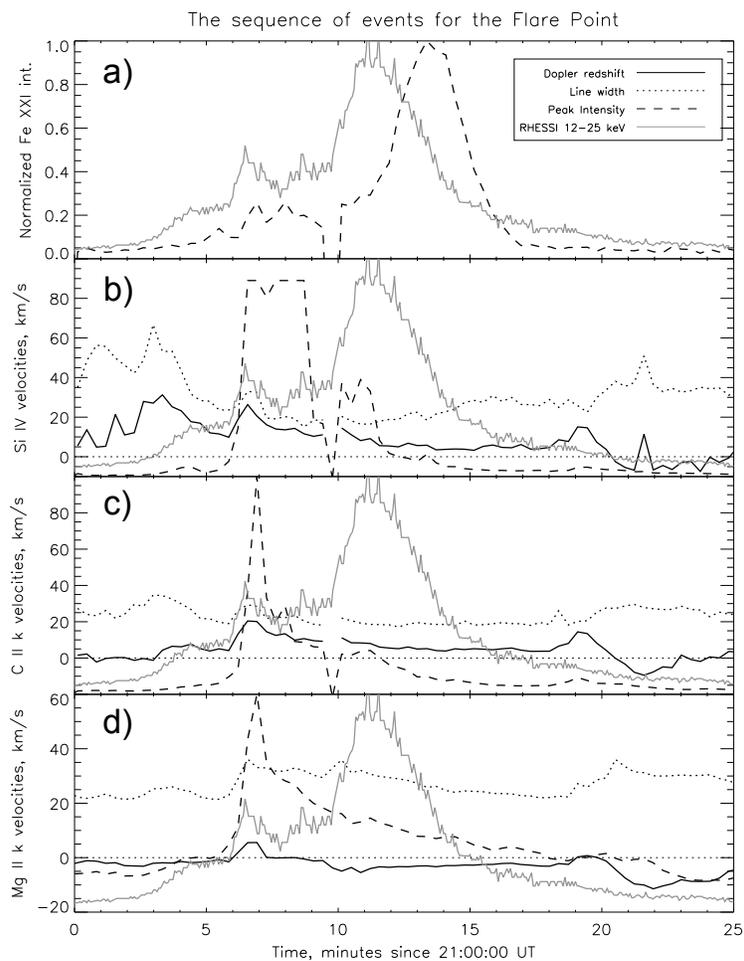}
\caption{Time evolution of the Doppler shift (black solid), line width (black dotted), and peak intensity (black dashed) for the flare event at the location marked with black cross and number 2 in Figure~\ref{figure3}: a)~Fe\,XXI (peak intensities only), b)~Si\,IV, c)~C\,II\,k, and d)~Mg\,II\,k lines. The 12-25\,keV X-ray light curve (gray solid) is overplotted in all panels.}
\label{figure8}
\end{figure}

\begin{figure}
\centering
\includegraphics[width=\linewidth]{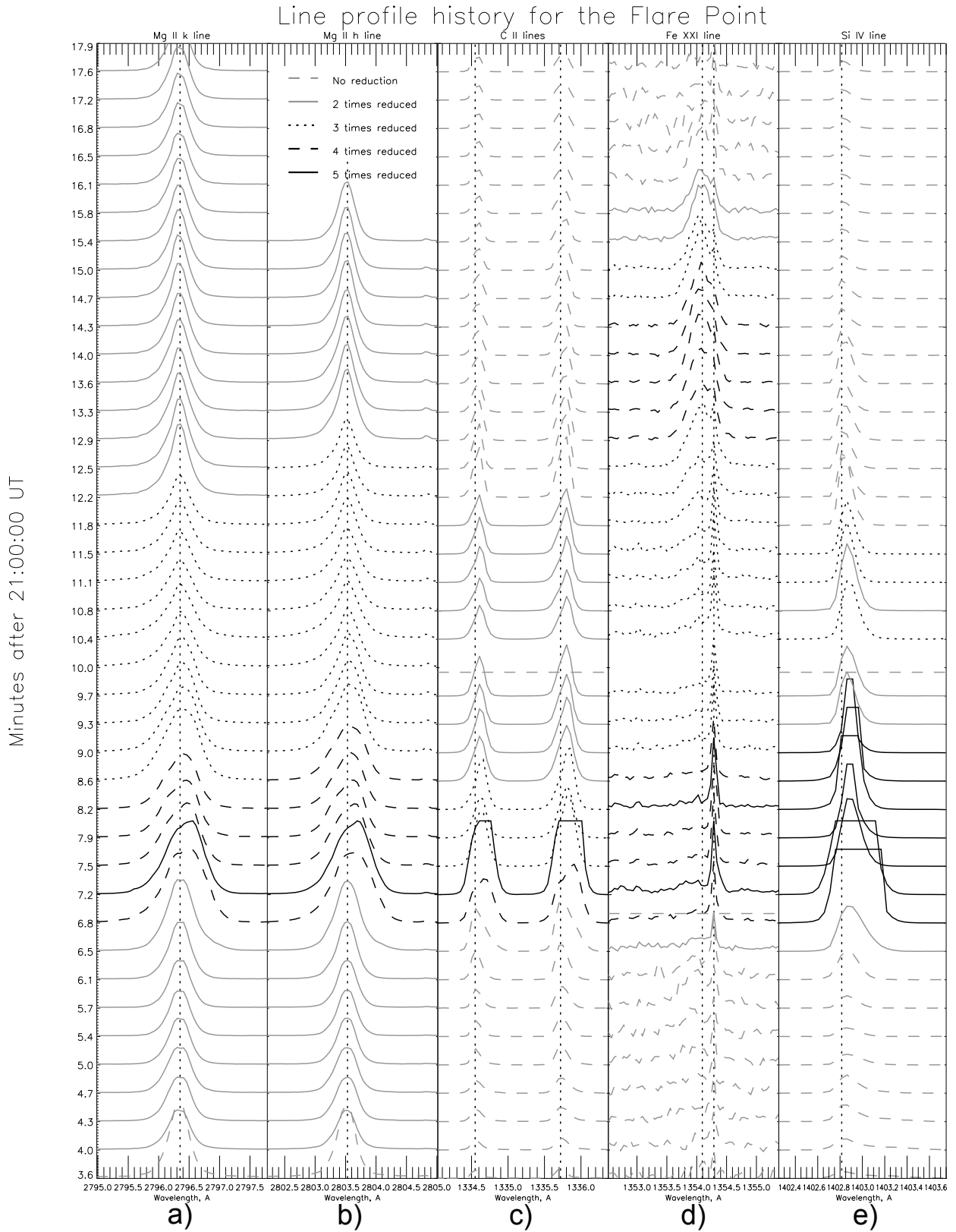}
\caption{Line profile evolution for the flare event at the location marked with the black cross and number 2 in Figure~\ref{figure3}for: a)~Mg\,II\,k, b)~Mg\,II\,h, c)~C\,II\,h\&k, d)~Fe\,XXI, and e)~Si\,IV lines. Grey scale line styles of the profiles correspond to the amplitude scaling factors (see text and the legend at the top of panel b) for details).}
\label{figure9}
\end{figure}

The Doppler shift estimates, line width estimates, peak intensities, and the evolution of line profiles are presented in Figures~\ref{figure8}~and~\ref{figure9} for the characteristic flare point. The RHESSI 12-25\,keV flux is also plotted in Figure~\ref{figure8}. All notations are the same as those in Figures~\ref{figure4}~and~\ref{figure5}. The reference wavelengths of the Mg\,II, C\,II and Si\,IV lines are obtained from the atomic spectral line database \citep{Kurucz95}, and additionally calibrated using quiet-Sun data near the studied active region. The time scale is shown relative to 21:00:00\,UT in minutes. We also consider the peak intensities and profiles for one more line: Fe\,XXI 1354.1\,{\AA}. This is a forbidden transition line that is formed in a hot (about $1.1\cdot 10^{7}$\,K) coronal plasma, and appears during the flare in the IRIS O\,I spectral window. The emission of this line starts at $\sim$21:10\,UT, and its peak occurs at $\sim$21:13\,UT. This is not the first observation of this line during solar flares. In a recent paper, \citet{Young14} observed this line during the X\,1 class flare that occurred on 2014 March 29 at 17:48\,UT, and studied its evolution and Doppler shifts. \citet{Tian14b} found even stronger redshift of this line in a C\,1.8 class flare on 2014 April 19, interpreted as a reconnection outflow, and also a strong blueshift presumably corresponding to the chromospheric evaporation. We adopt the same interpretation in our paper.

It is important to note that the Fe\,XXI 1354.1\,{\AA} line is blended: the main blend of this line is C\,I 1354.29\,{\AA} line in the red wing, and other blends are Fe\,II 1354.013{\AA} and Si\,II 1353.718{\AA} lines in the blue wing (see Figure~2 of \citet{Tian14b}, and the Appendix A of \citet{Young14} for details). Therefore, it is necessary to separate contributions of these lines to obtain the Doppler redshift and line-width estimates. The center of gravity definition is not applicable in this case. In this paper, we consider only two main characteristics of the Fe\,XXI line: the line peak intensity and the wavelength corresponding to this peak. The C\,I 1354.29\,{\AA} line, however, helps us to calibrate the reference wavelength for the Fe\,XXI line. The C\,I line appears as a very narrow and sharp peak, which means that this line is formed in deep layers of the solar chromosphere. This line does not show any significant Doppler shifts: we calibrated the reference wavelength of this line at several points of the flare structure and did not find any significant deviations of the line peak wavelength from the reference value. We shifted the IRIS wavelength vector in the IRIS O\,I spectral window such that the reference wavelengths of the C\,I and Fe\,XXI lines correspond to the tabulated values: 1354.094\,{\AA} and 1354.288\,{\AA} \citep{Vilhu01}. Of course, the other two lines (Fe\,II 1354.013{\AA} and Si\,II 1353.718{\AA}) have an impact on the resulting Fe\,XXI 1354.1\,{\AA} line profile. However, as we study the dynamics of the hot plasma, the broad Fe\,XXI 1354.1\,{\AA} line obviously becomes dominant one in the O\,I spectral window (see behaviour of this line for the characteristic flare point in Figure~\ref{figure9}). Therefore, in this study we neglect the impact of these blends on the Fe\,XXI line.

Figure~\ref{figure8} shows that emission of the chromospheric and transition region EUV lines significantly grows during the flare, and reaches maximum at $\sim$21:07\,UT. This maximum corresponds to the small peak of flux in the 12-25\,keV X-ray energy band. The 6-12\,keV signal also shows increase at the same time; however, it is less sharp than the 12-25\,keV signal. The intensities of the IRIS lines have a clearly observed phase of an exponential decay. The line width also grows during the flare by a factor of $\sim$1.5, but this change is not so significant as the increase of the line peak intensity. If the broadening nature is thermal, the temperature of the region where lines are effectively formed increases just by a factor of 2. The line intensities probably grow due to an increase of the number densities of the Mg$^{+}$, C$^{+}$, and Si$^{3+}$ ions. This increase may happen due to a sharp increase of ionisation caused by nonthermal mechanisms, e.g., by nonthermal electrons accelerated in the reconnection region. This could also explain the correspondence between the 12-25\,keV X-ray flux peak and emission peaks of lines.

The peak of the Fe\,XXI line emission is delayed relatively to the peak emission of the other lines by about 6 minutes. The peaks of the Mg\,II\,h\&k, Si\,IV, and C\,II lines, shown in Figure~\ref{figure9}, are slightly redshifted during their strongest emission phase, and the emission of the Fe\,XXI line becomes obvious at $\sim$21:10\,UT. Perhaps, the delay of the Fe\,XXI intensity relative to the cool lines is because the maximum of the Fe\,XXI intensity is associated with postflare loops while the maximum intensities of the cool lines are related to chromospheric heating mainly in the impulsive phase.

\begin{figure}
\centering
\includegraphics[width=11cm]{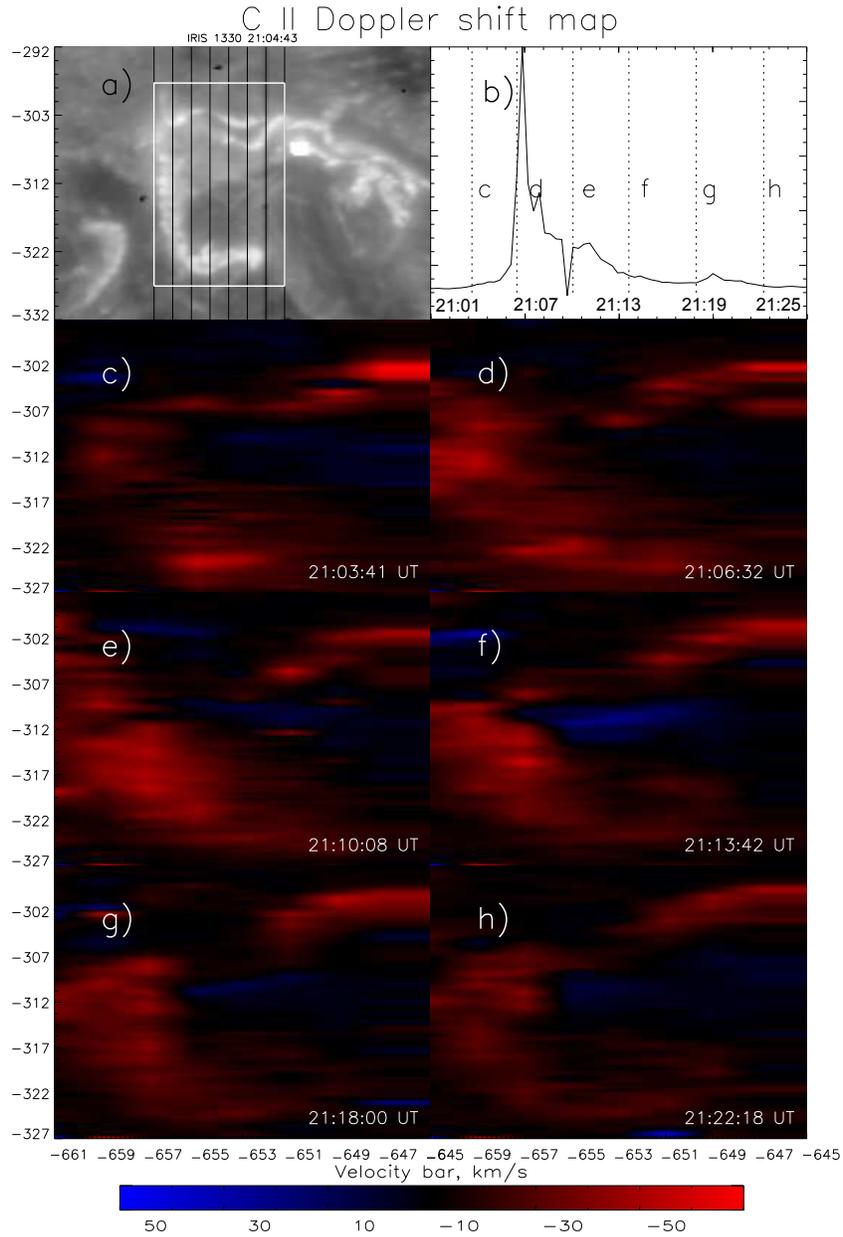}
\caption{Evolution of the Doppler shift map for the C\,II\,k line. Description of images: a)~flare region observed in the 1330\,{\AA} line at 21:04:43\,UT with the marked IRIS slit positions (vertical black lines). The white rectangle corresponds to the region, for which the Doppler shift maps are plotted in panels c-h) for six moments of time shown in the bottom right corners, and marked by dashed lines in panel b); b) peak intensity profile for the C\,II\,k line as a function of time. Red color corresponds to the redshifted regions, and blue color correspond to the blueshifted regions.}
\label{figure10}
\end{figure}

Figure~\ref{figure10} displays a sequence of the Doppler shift maps for the C\,II\,k line. The maps represent the Doppler shift estimates determined at the 8 slit positions with a 0.33 arcsec resolution along the slits, and linearly interpolated between the slit positions. The selected time moments are marked with dashed lines crossing the peak intensity plot for the C\,II line. It is clear that most of the flaring region observed in the 1330\,{\AA} IRIS image is redshifted. The redshifts at the selected flare point (Figure~\ref{figure8}) are around 20\,km/s for the transition region C\,II and Si\,IV lines, and are near zero for the chromospheric Mg\,II\,k line. Redshifts across the flaring region are greater than or around 50\,km/s. It is interesting that these redshifts exist before the impulsive phase of the flare (see Figure~\ref{figure10}c) and after the impulsive phase (see Figure~\ref{figure10}h), and do not change their spatial extension, in general.

According to the standard flare model \citep[e.g.][]{Sturrock68}, redshift for the chromospheric and transition region lines and blueshift for the hot coronal lines correspond to the chromospheric evaporation process. Figure~\ref{figure11} displays the Doppler shift evolution for the Fe\,XXI line. The Doppler velocity estimate for the Fe~XXI line is calculated as the difference between the wavelength of the line peak and the reference wavelength for the Fe\,XXI line, measured in km/s. Reconstructed from the RHESSI data, the 12-25\,keV X-ray sources are overplotted for the corresponding time intervals. The six displayed maps are prepared in the same way as the Doppler shift maps for the C\,II\,k line, and correspond to the same time moments. Quite strong blueshifts (around -50\,km/s) are observed for the Fe\,XXI line across the region. \citet{Young14} also observed very strong blueshifts of 100-200\,km/s for the Fe\,XXI line for a different event.

Figure~\ref{figure12} shows the composition of normalized integrated characteristics of the event: the C\,II\,k and Fe\,XXI integrated intensities, GOES temperature and emission measure, and the RHESSI 12-25\,keV light curve. The C\,II\,k and Fe\,XXI intensities represent the peak intensities of these lines integrated over the flare region indicated in Figure~\ref{figure10}a by the white rectangle. The GOES temperature and emission measure are constructed with the same algorithm as in Figure~\ref{figure2}. Notice the correspondence between the first two C\,II\,k integrated intensity peaks and two small peaks of the X-ray emission in the energy range 12-25\,keV, the correspondence between the Fe\,XXI integrated intensity and the emission measure peaks and growth phases, and also the 6\,min delay of the Fe\,XXI intensity peak relative to the C\,II\,k intensity peak.

Figure~\ref{figure13} displays the line peak intensity map for the Fe\,XXI line for different time moments. It is constructed in the same way as the images in Figures~\ref{figure10} and~\ref{figure11}. This figure represents emission of the hot $1.1\cdot 10^{7}$\,K coronal plasma. It shows that the strongest emission of the Fe\,XXI line occurs not only above bright chromospheric structures observed in the 1330\,{\AA} line, but also in the middle of the region. The source of the strongest emission is not stationary; it moves across the region in time. The Fe\,XXI line becomes very intensive across the flare region at approximately 21:12\,UT. In addition, in Figure~\ref{figure14} we present the corresponding images of the EUV 131\,{\AA} emission from the SDO Atmospheric Imaging Assembly (AIA,~\citet{Lemen12}). The X-ray 6-12\,keV and 12-25\,keV sources are overplotted for the corresponding time intervals. The Fe\,XXI intensity maps are similar to the AIA images. This is because the AIA 131\,{\AA} response function has one peak at $1-2\cdot 10^{7}$\,K, and is sensitive to Fe\,XX, Fe\,XXI and Fe\,XXIII lines at these temperatures \citep{Schmelz13}. This fact was first discussed in the paper of \citet{Young14}. However, because the 131\,{\AA} signal is overexposed during this and other flares, the information about the location and structure of the hot plasma sources is lost. The Fe\,XXI intensity maps help to resolve these sources and understand the location of the hot evaporated plasma and its expansion in the corona.

\subsection{Analysis of RHESSI spectra}

It is important to understand the energy sources and mechanisms of the energy transport which caused the observed chromospheric evaporation. To understand this we analyze the X-ray spectra obtained by RHESSI. In Figure~\ref{figure15} we present results of our spectral analysis of the X-ray data for three time intervals (three columns in Figure) corresponding to the integration intervals of the RHESSI images shown in Figure~\ref{figure12} as light-green stripes. The time intervals and model parameters are shown in Table~1.

For the X-ray spectral analysis of the first time interval (21:03:20-21:05:24\,UT, left column in Figure~\ref{figure15}) we consider three different models: pure thermal and nonthermal models, and a combination of thermal and nonthermal models. For the pure thermal model, we obtained good fitting results with $\chi^2=2.76$ and reasonable values of $T\approx{}24.9\cdot{}10^{7}\,$K and $EM\approx{}1.9\cdot{}10^{46}\,$cm$^{-3}$. The spectral fitting using the pure nonthermal model \citep[thick-target model of][]{Brown71} gives nonphysically large fluxes of nonthermal electrons $\sim 10^{37}-10^{38}$ electrons/s with the low-energy cutoff of $\lesssim$8 keV. The combined model did not provide a stable fitting result for the first time interval corresponding to a preheating phase. Thus, we consider the pure thermal model as the best-fit model of the first spectrum.

The X-ray spectra for the second (21:07:44-21:09:24\,UT) and the third (21:11:04-21:12:28\,UT) time intervals can be explained using two approaches: 1) a superposition of a single-temperature model with the thick-target nonthermal model; 2) a double-temperature model. The first fitting has six free parameters: temperature $T$, emission measure $EM$, nonthermal electrons flux $F(E>E_{low})$ above low-energy cutoff $E_{low}$, spectral power-law index $\delta$ of nonthermal electrons, and high energy cutoff $E_{high}$. In the double-temperature model we vary four parameters: temperature $T_1$ and emission measure $EM_1$ for the ``warm'' plasma component, and $T_2$ and $EM_2$ for the ``hot'' component. The fitting results are presented in Table~1. Both approaches show quite reasonable fitting with $\chi^2<3$. Spatially resolved X-ray observations by RHESSI reveal that the 6-12 keV and 12-25 keV emission sources coincide with each other for these two time intervals (this is clearly seen in Figure~\ref{figure11}e,f). In terms of the thermal-plus-nonthermal model this means that we observe an instantaneous plasma heating by nonthermal electrons, which also generate thick-target bremsstrahlung from the same coronal volume (coronal thick-target emission, \citet{Veronig04}). However, the double-temperature model is also possible due to the unresolved fine structures of the X-ray sources. At this point, we cannot unambiguously determine which of the models gives a more realistic explanation.

\subsection{Summary of Observations}

In this subsection we summarize briefly the main findings:
\begin{itemize}
  \item A strongly redshifted jet appears near the bright loop-like structure just prior the flare. The jet starts at $\sim$20:48:00\,UT and lasts until $\sim$21:03:30\,UT. In the line profiles (see Figure~\ref{figure5}), this jet appears as an additional redshifted (for $\sim$100\,km/s) peak in the Mg\,II lines and C\,II lines.
  \item A sharp increase of the C\,II\,k line intensity across the flaring region (see Figure~\ref{figure7}) and in the characteristic flare point near a X-ray footpoint (see Figure~\ref{figure8}). The C\,II\,k emission points are localized (with $\sim$1\,arcsec spatial size), and occur at various points prior to the X-ray peak of the flare. In the characteristic flare point, a sharp increase of emission of the C\,II, Mg\,II, and Si\,IV lines is not accompanied by a significant increase of the line widths, indicating that the increased emission is probably associated with nonthermal processes.
  \item A strong stable redshift ($\sim$50\,km/s) in the C\,II\,k line is observed across the flaring region. The redshift is observed before the impulsive phase of the flare, as well as in the decaying phase of the flare (see Figures~\ref{figure10}c-h), and its spatial configuration does not change significantly during the impulsive phase of the flare.
  \item A blueshift of $\sim$50\,km/s across the flaring region is observed in the forbidden Fe\,XXI line emitted by the high-temperature plasma. Initially, the blueshifts are in two separate regions (Figure~\ref{figure11}d). Later, they fill the whole flare region (Figures~\ref{figure11}e,f). These observations correspond to the phenomenon called ``chromospheric evaporation''~--- the upflow of the chromospheric plasma heated by a beam of accelerated electrons or by other mechanisms, which we discuss in the following section.
\end{itemize}

\section{Discussion}

Historically, two types of the chromospheric evaporation are distinguished. The first type, so-called ``gentle'' evaporation \citep{Antiochos78}, can be driven by electron heat conduction, and corresponds to relatively low subsonic velocities of the evaporating plasma ($v$\,$<$\,200\,km/s) and long time scales ($\sim{}$minutes), according to \citet{Zarro88}. The second type, so-called ``explosive'' evaporation, occurs when the upper chromosphere is overheated by a high energy flux of accelerated particles, quickly reaches coronal temperatures, and therefore expands. This type of evaporation is characterized by supersonic upflows of evaporating plasma ($v$\,$>$\,200\,km/s) and relatively short time scales \citep{Zarro88}.

According to the hydrodynamic simulations of electron heating \citep[e.g.][]{Fisher85a,Fisher85b,Fisher85c}, the gentle plasma evaporation can be caused not only by electron heat flux but also by a relatively low flux of accelerated electrons ($<10^{10}$\,ergs\,cm$^{-2}$\,s$^{-1}$). In this case the evaporation flow results in Doppler blueshifts of several tens of km/s of spectral lines formed in the upper chromosphere and transition region. Evidence of this kind of evaporation was demonstrated by \citet{Milligan06}. The explosive evaporation corresponds to higher energy fluxes ($>5\cdot10^{10}$\,ergs\,cm$^{-2}$\,s$^{-1}$) and, in addition to the supersonic upflows of evaporated plasma observed as the blueshift of hot coronal lines, results in redshifts of tens of km/s of cooler chromospheric and transition region lines. The simulations of the explosive heating in the radiative-hydrodynamic thick-target model \citep{Kostiuk75} also showed the presence of a temperature region that divides the atmosphere into redshifted and blueshifted parts \citep[e.g.][]{Livshits83,Kosovichev86,Liu09} with the division temperature about $10^{6}$\,K. A blueshift of hot coronal lines and a redshift of cooler lines was observed in many studies \citep{Doschek13,Milligan09,Raftery09,Brosius04,Brosius07}. These observations qualitatively support the standard thick-target model with the explosive evaporation.

\begin{figure}[t]
\centering
\includegraphics[width=0.7\linewidth]{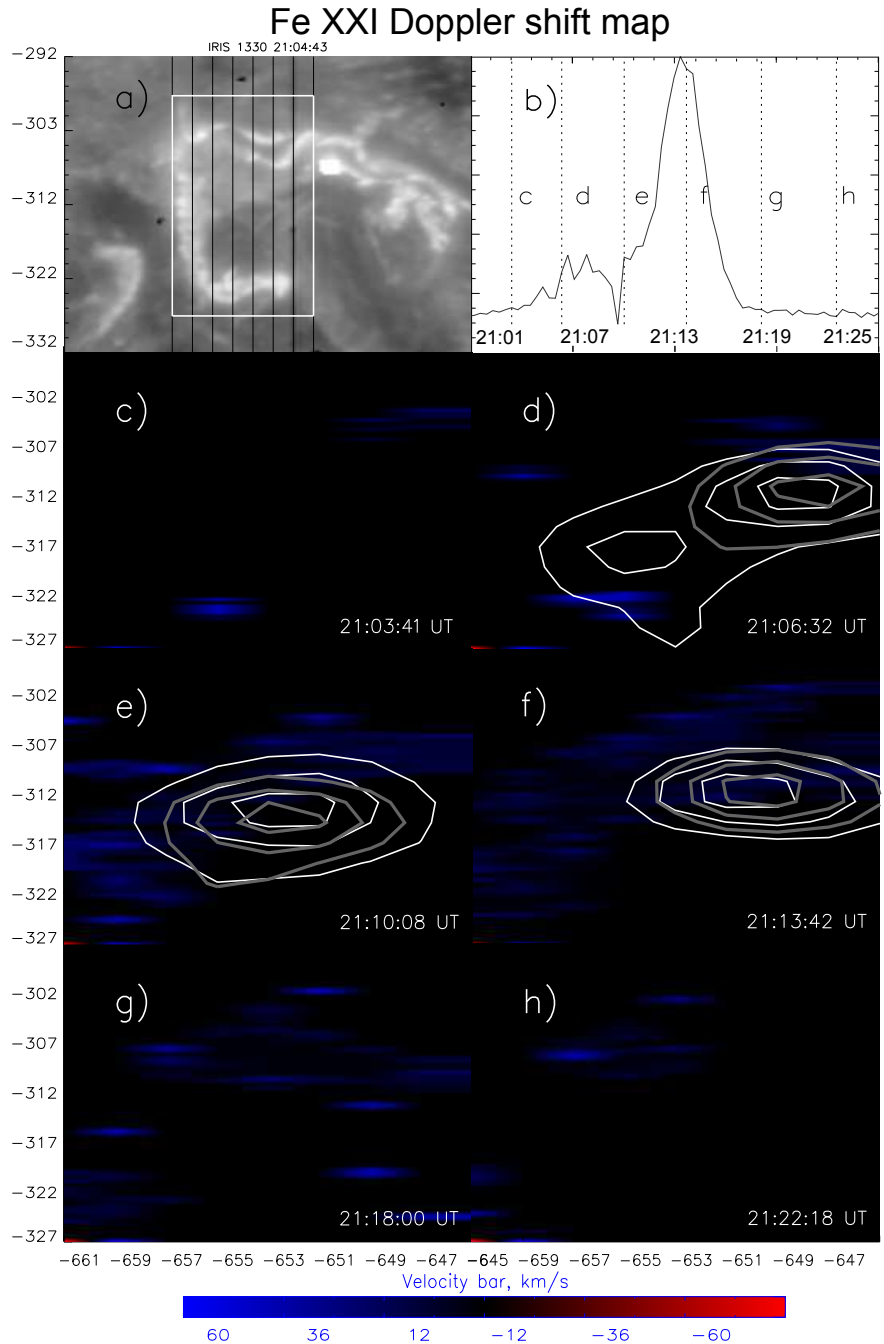}
\caption{Doppler shift maps for the Fe\,XXI line for six moments of time corresponding to the images in Figure~\ref{figure10}c-h. These images represent the Doppler shift maps for the same white box as in Fig.~\ref{figure10}a). The Doppler shift range is from -50\,km/s to 50\,km/s. The corresponding 12-25\,keV (white) and 6-12\,keV (grey) X-Ray sources for d)~21:04:00-21:06:00\,UT, e)~21:07:50-21:09:50\,UT and f)~21:12:00-21:14:00\,UT are overplotted. Panel b) displays the peak intensity profile for the Fe\,XXI line as a function of time.}
\label{figure11}
\end{figure}

\begin{figure}[t]
\centering
\includegraphics[width=0.9\linewidth]{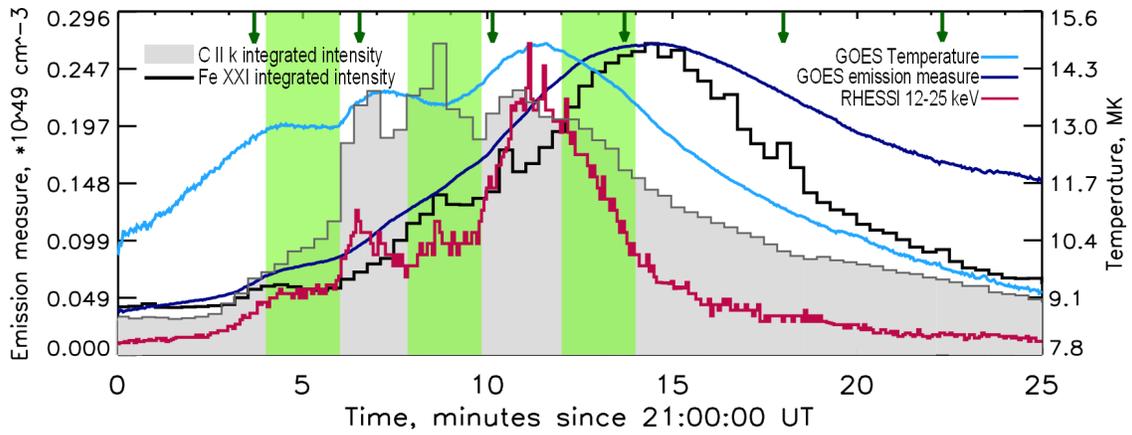}
\caption{Integrated intensities of the C\,II\,k (gray) and Fe\,XXI (black) lines, the GOES temperature and emission measure, and 12-25\,keV light curve from RHESSI during the flare main phase. Light-green stripes indicate the time intervals for which the 12-25\,keV and 6-12\,keV X-ray sources were reconstructed from the RHESSI data. Green arrows correspond to the time moments of the Doppler shift maps in Figures~\ref{figure10}c-h~and~\ref{figure11}c-h}
\label{figure12}
\end{figure}

\begin{figure}
\centering
\includegraphics[width=0.7\linewidth]{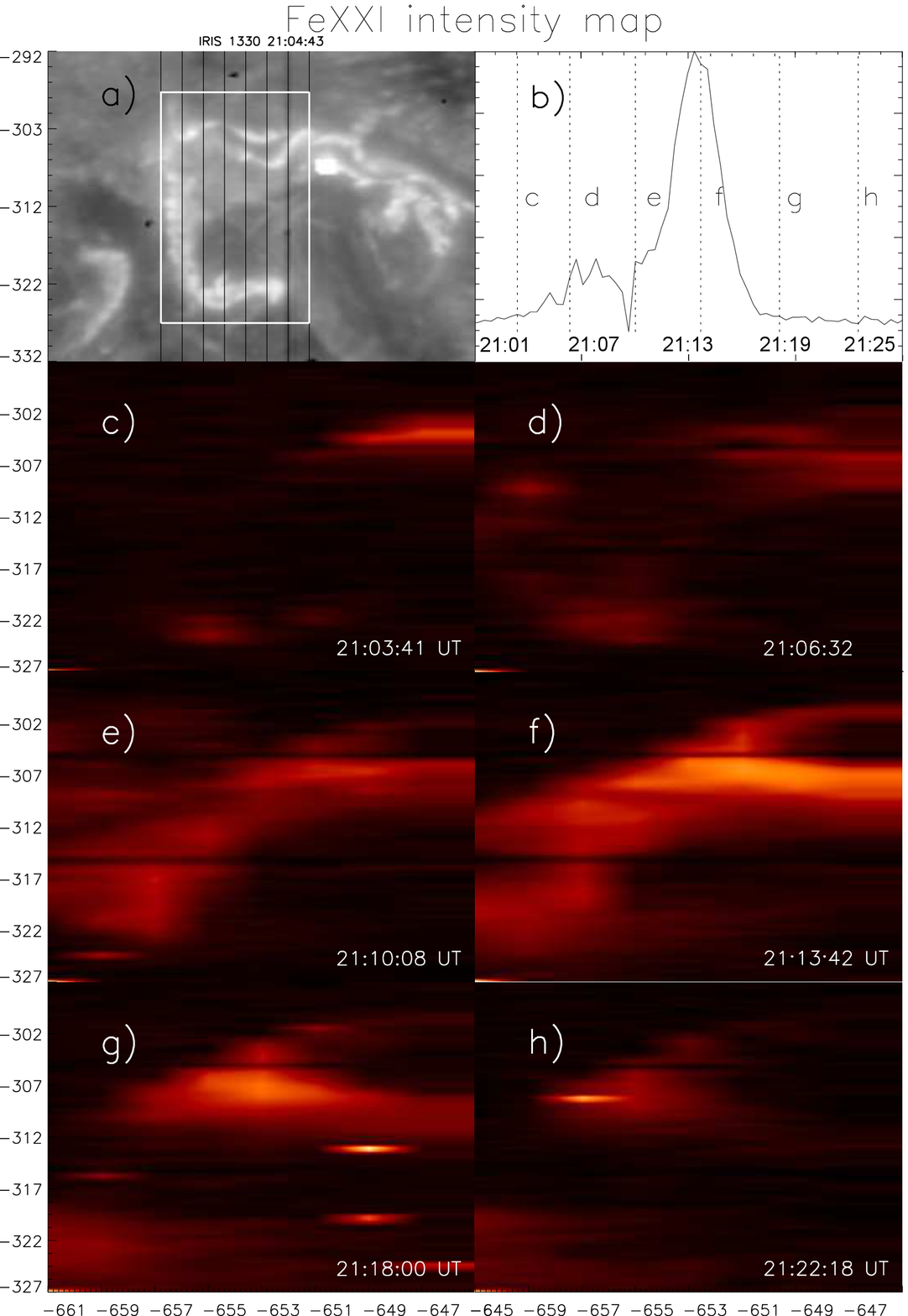}
\caption{Evolution of the Fe\,XXI line intensity. Description of images: a)~flare region observed in the 1330\,{\AA} line at 21:04:43\,UT with the marked IRIS slit positions (vertical black lines). The white rectangle corresponds to the region, for which the line intensity maps are plotted in panels c-h) for six moments of time shown in the bottom right corners, and marked by dashed lines in panel b); b) peak intensity profile for the Fe\,XXI line as a function of time. The images are plotted in the logarithmic scale.}
\label{figure13}
\end{figure}

\begin{figure}[t]
\centering
\includegraphics[width=0.7\linewidth]{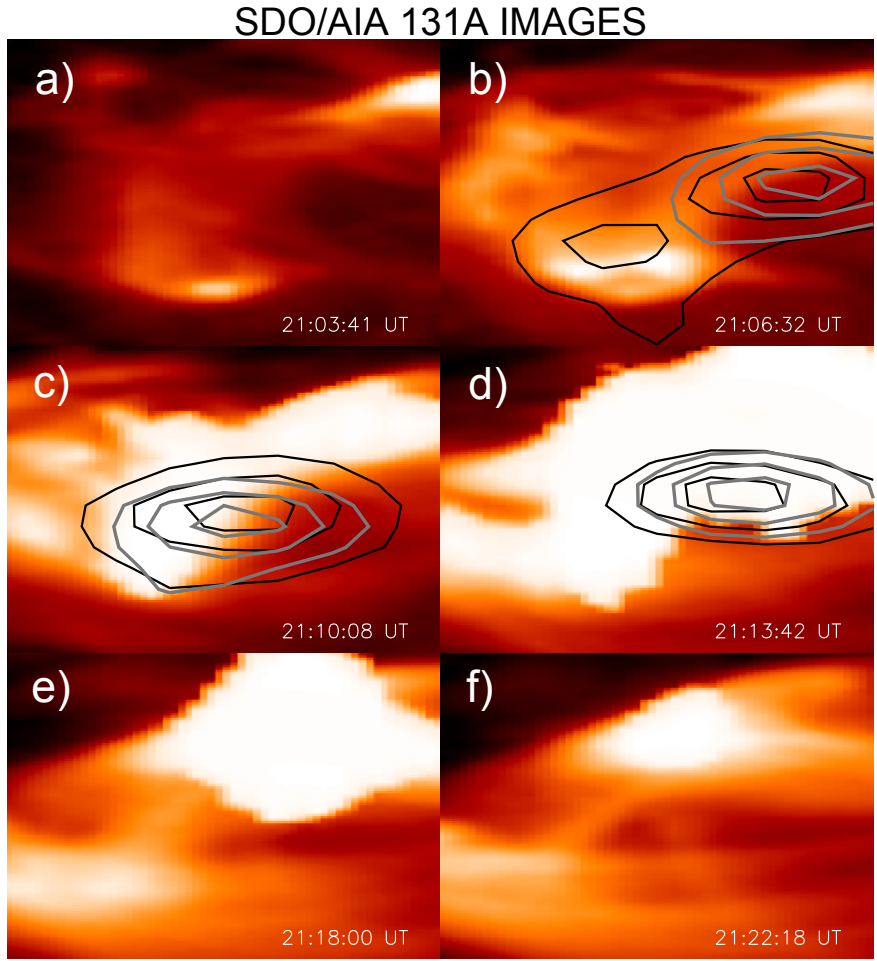}
\caption{The 131\,{\AA} EUV images from SDO/AIA for six moments of time corresponding to the IRIS data shown in panels c)-h) of Figure~\ref{figure13}. The images are plotted in the logarithmic scale. The 12-25\,keV (black) and 6-12\,keV (grey) X-ray sources reconstructed from the RHESSI data for the three corresponding intervals: b)~21:04:00-21:06:00\,UT, c)~21:07:50-21:09:50\,UT and d)~21:12:00-21:14:00\,UT are overplotted.}
\label{figure14}
\end{figure}

\begin{figure}[t]
\centering
\includegraphics[width=\linewidth]{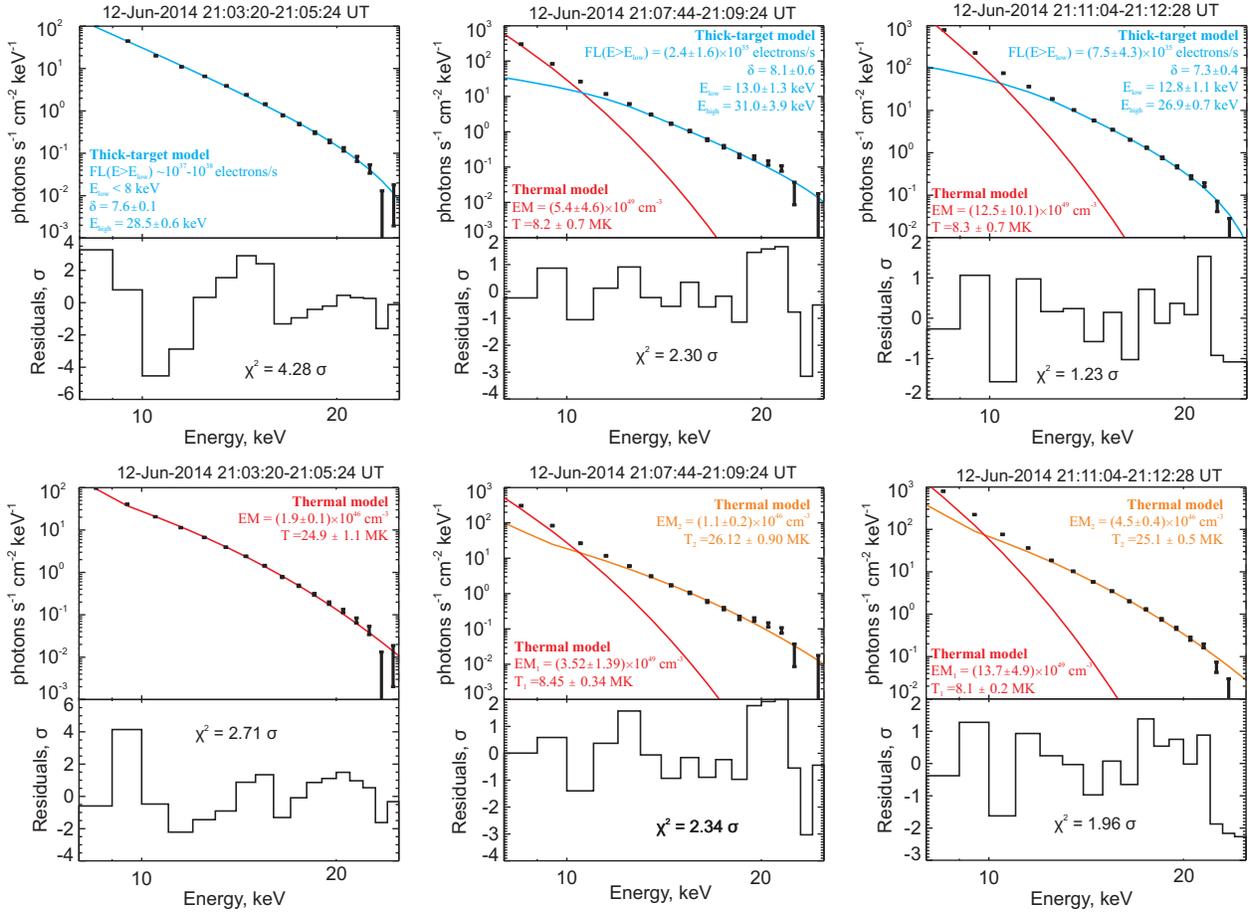}
\caption{The RHESSI X-ray spectra and the best-fit models for 21:03:20-21:05:24\,UT (left column), 21:07:44-21:09:24\,UT (central column) and 21:11:04-21:12:28\,UT (right column) time intervals.}
\label{figure15}
\newpage
\end{figure}

The most recent observations of the chromospheric evaporation made by IRIS were presented by \citet{Tian14b}. The blueshifts of $\sim$260\,km/s of the hot Fe\,XXI line were observed simultaneously with the significant redshifts of $\sim$50\,km/s of the upper chromospheric and transition region O\,IV, Si\,IV, C\,II and Mg\,II lines. Thus, these data may correspond to the explosive chromospheric evaporation. However, it is not easy to unambiguously determine the type of evaporation in the flare considered in our paper. On one hand, the theoretical prediction of the division temperature in the explosive type of evaporation is supported by our results: significant redshifts (about 50\,km/s) are observed for the transition region C\,II\,k line formed at $10^{4}-2\cdot 10^{4}$\,K across the flaring region, and the blueshift of $\sim$50\,km/s is observed for the $1.1\cdot 10^{7}$\,K coronal Fe\,XXI line. The strongest shifts illustrated in Figures~\ref{figure10}d~and~\ref{figure11}d for 21:08:00\,UT were observed during the impulsive phase of the flare from $\sim$21:03:00\,UT to $\sim$21:11:00\,UT (Figure~\ref{figure2}). On the other hand, velocities of the evaporating plasma in our case are definitely subsonic. The Doppler shift velocities are about 50\,km/s for the Fe\,XXI line. Also, the characteristic time scales of the evaporation process are rather long: the red and blue shifts are significant not only during the flare impulsive phase, but also at 21:11:33\,UT and 21:15:08\,UT (Figures~\ref{figure10}e,f,~\ref{figure11}e,f), definitely after the impulsive phase. Thus, it is likely that we observe the gentle chromospheric evaporation. In this case, the origin of the redshifts during the evaporation is unclear. However, these redshifts definitely existed before and after the impulsive phase of the flare (we observe redshifts even at 21:03:41\,UT and 21:22:18\,UT as shown in Figures~\ref{figure11}c~and~\ref{figure11}h), and these are unlikely connected with the chromospheric evaporation process. Probably, these spatially broad redshifts are related to the general dynamics of the active region and caused by a cooling plasma falling down along the magnetic loops. As discussed above, \citet{Milligan06} observed weak upflows in cool He\,I and O\,V lines for a gentle evaporation case. The question why in this case we do not observe blueshifts in the cool C\,II\,k line is still open. The most possible reason is that the expected blueshifts are relatively weak and are not seen because of the large and intense redshifts of the background plasma. Also, the estimated blueshifts of cool lines made by \citet{Milligan06} are low and with large uncertainties (13$\pm$16 and 16$\pm$18 for He\,I and O\,V lines). Thus, these line shifts are not easily detectable because the spectral resolution of IRIS in this range is 26\,m{\AA}, which corresponds to $\sim$6\,km/s resolution of the line-of-sight velocity, just as twice as low the expected blueshifts.

The first signatures of the evaporation are observed as two local maximums of the Fe\,XXI blueshifts (Fig.~\ref{figure11}d). The AIA\,131\,{\AA} image in Fig.~\ref{figure14}c shows that these zones are connected by a coronal loop system. It is possible that the hot evaporated plasma fills the loop system, and that we observe this effect in Fig.~\ref{figure11}e-f. Let us compare the amount of the hot plasma (the total number of electrons), which could be evaporated from the chromosphere into the corona: $N_{ev} \sim 2 n_{chr}\cdot{}v_{chr}\cdot{}S\cdot{}\delta{}t$ with the amount of the hot plasma emitting in the X-ray range: $N_{X-ray} \sim \sqrt{EM\cdot{}S\cdot{}L}$. Here, $n_{chr} \sim 10^{10}$ cm$^{-3}$ is the number density of electrons in the lower corona or upper chromosphere from where the evaporation occurs; $v_{chr} \approx 50$~km~s$^{-1}$ is the speed of the evaporated plasma estimated from the observed blueshifts of the Fe\,XXI line; $S \sim 10^{17}$~cm$^{2}$ is the area of the Fe\,XXI sources in Figure~\ref{figure11}d; $\delta{}t \sim 10^{2}$\,s is a characteristic time scale of the evaporation process; $EM \sim 10^{46}$~cm$^{-3}$ is the emission measure of the X-ray source, estimated from the spectral analysis of the RHESSI X-ray observations; $S \sim 10^{17}$~cm$^{2}$ is the area of the X-ray source, and $L \sim 10^{9}$~cm is the source length. Using these parameters, we find that $N_{ev} \sim N_{Xray} \sim 10^{36}$ electrons. Although our estimates are quite rough, nevertheless they give us an evidence that the hot X-ray emitting plasma of the flare is initially evaporated from the chromosphere. The estimation of time, which is needed for the evaporated plasma to fill the loop, is $t_{fill}\sim{}L/v_{chr}\sim200$sec, which is consistent with the 3\,min delay of the Fe\,XXI emission peak relative to the 12-25\,keV X-ray peak (see Figure~\ref{figure12}).

Analysis of the X-ray spectra shows that the accelerated electrons have a very soft spectrum with the power-law index reaching values more than $7$. This indicates that the large part of the accelerated electrons is suprathermal. Moreover, it is possible to consider the thermal interpretation of the whole X-ray spectrum with a double-temperature model (Sec.3.3). The chromospheric evaporation can be stimulated not only by an injection of the accelerated nonthermal particles to the chromosphere from the corona. The heat flow from a primary energy release site may be another mechanism to overheat the chromosphere and induce a chromospheric evaporation \citep{Antiochos78}. This scenario is especially viable during the preheating phase, as the X-ray spectrum can be explained by a single-temperature model without significant fluxes of nonthermal electrons. One can make some simple estimations of the energy deposited by the heat flux and by the nonthermal electrons into the dense solar atmosphere. To estimate the heat conduction flux, we use the classical expression \citep{Spitzer53}:
\vskip-.6cm
\begin{eqnarray}
    Q_{cl}=k_{0}T_{e}^{5/2}S\nabla{}T_e\sim{}\frac{k_{0}T_{e}^{7/2}S}{L}, \nonumber
\end{eqnarray}
Where $k_0\approx$10$^{-6}$\,erg\,s$^{-1}$\,cm$^{-1}$\,K$^{-7/2}$, $L$ is a length scale of the temperature gradient, $S$ is the cross section of the magnetic loops. Calculation gives $Q_{cl}\sim 10^{28}$\,erg/s for the $T_e\approx 25$\,MK, $S\sim10^{17}$\,cm$^2$, and $L\sim 10^9$\,cm, estimated from the observations.

The power, $P_{nonth}$, of the accelerated electrons can be calculated from:
\vskip-.6cm
\begin{eqnarray}
    P_{nonth}=\frac{\delta-1}{\delta-2}F(E>E_{low})E_{low} \nonumber
\end{eqnarray}
Substituting the parameters obtained in Sec.3.3: $F=2.4\times10^{35}$\,electrons/s, $E_{low}=13$\,keV and $(\delta-1)/(\delta-2)\sim 1$ for $\delta=8.1$, we get: $P_{nonth}\approx 5\times 10^{27}$\,erg/s. We see that $P_{nonth}$ is of the same order of magnitude as $Q_{cl}$. Thus, both the accelerated electrons and the heat conduction flux can be responsible for the observed chromospheric evaporation. Interestingly, if we calculate the electron energy flux, $F/S$, we find that it is equal to $5\times10^{10}$\,ergs\,cm$^{-2}$\,s$^{-1}$ for $S\sim10^{17}$\,cm$^2$. According to the simulations of \citet{Fisher85a}, this flux is comparable with the minimal flux required for the explosive type of evaporation ($\approx{}5\times 10^{10}$\,ergs\,cm$^{-2}$\,s$^{-1}$). However, because of the subsonic velocities the plasma evaporation process in our event is classified as the ``gentle'' evaporation. One of the possibilities to resolve this contradiction is to assume that the calculated flux of nonthermal electrons is slightly overestimated. Another possibility is that the real physical conditions of the lower solar atmosphere in the flare region differ from those used by \citet{Fisher85a} in their calculations.

It is important to note that the flare region is approximately located at $\sim$50 degrees from the disc center (the angular disc coordinates are (-650$^{\prime\prime}$,-310$^{\prime\prime}$) relative to the disc center). Therefore, the estimated velocities and spatial sizes are affected by projection effects. For example, if the plasma is moving radially then its velocity is $\sim$1.5 times greater than the estimated line-of-sight velocity. However, the projection effects do not change our principle results. Similar blueshifts/redshifts of the IRIS lines are observed across the flaring region, so that the projection effects due to variation in the flow topology do not affect the conclusion about the nature of flows in this event.

The 12-25\,keV X-ray sources reconstructed using the RHESSI data (for example, shown in Figure~\ref{figure11}d) show a characteristic two-footpoint structure. However, these double X-ray sources are more likely to be located in the corona rather than in the chromosphere. Using the coronal thick-target model \citep{Veronig04}, we estimate whether the energetic electrons can lose all their energy in the corona before reaching the chromosphere. According to \citet{Brown73} and \citet{Veronig04}, the energy of electrons collisionally stopped in the column density $N$, is $E_{loop}\sim{}9N_{19}^{1/2}$\,(keV), where $N_{19}=N/10^{19}$\,(cm$^{-2}$). We can estimate $N$ as $N\sim{}N_{X-ray}/S \sim \sqrt{EM\cdot{}L/S}$. Using the X-ray source parameters, we obtain $E_{loop}\sim{}28$\,keV. As noticed before from the RHESSI spectral analysis, the maximum energy of the accelerated electrons is $\sim{}30$\,keV. Thus, the bulk of the accelerated electrons is possibly stopped in the corona, while only a relatively small number of the most energetic electrons (with E~$\gtrsim$~30\,keV) from the non-thermal tail could penetrate into the chromosphere. Our suggestion is that these high-energy electrons are probably a reason of the enhanced C\,II\,k emission discussed above. In this case, modeling and analysis of the nonthermal emission of the chromospheric and transition region lines may potentially provide a new sensitive diagnostics of nonthermal particles. Of course, this suggestion requires further quantitative studies.

\section{Conclusion}

We have presented results of observations of the M1.0 class flare event occurred on 12 June, 2014 (SOL2014-06-12T21:12). The IRIS mission provided us with a great opportunity to investigate the spectral variations of different points of the flaring region over time, as well as changes of the flare spatial configuration. IRIS observed this event in the coarse raster mode and obtained high-resolution UV spectra and Slit-Jaw (SJ) images for almost the whole flaring region.

Several interesting findings can be highlighted from our analysis of this flare. First, we have observed a strongly redshifted plasma flow in the upper chromosphere just before the flare. The origin of this jet-like structure is unclear to us, and we plan to study its origin using simultaneous NST observations in our following paper. Second, we have observed the emission of the C\,II\,k line intensity in multiple points of the flaring region. We argue that this emission is probably due to excitation by nonthermal particles. Third, we observe the strong stable redshifts ($\sim$50\,km/s) almost in the entire flaring region in the transition region C\,II\,k line. These redshifts existed before, during and after the impulsive phase of the flare, and their spatial configuration did not change significantly. Thus, these redshifts are not related to the chromospheric evaporation process caused by this flare. A possible explanation is that the continuously heated and slowly evaporating plasma is falling back along the magnetic loops. Finally, we discuss in detail the spatial structure of the chromospheric evaporation process, observed as a blueshift in the Fe\,XXI line across the flaring region. The blueshifts of $\sim$50\,km/s and characteristic times of several minutes allow us to classify the observed dynamics as the ``gentle evaporation'' of the hot chromospheric plasma. The analysis of the X-ray spectra from RHESSI showed that this evaporation can be driven either by the heat conduction or by the heating by nonthermal accelerated particles. However, the pure thermal model requires a two-temperature distribution of the flare-heated plasma. While, in principle, this is possible due to unresolved fine structures, the mechanisms of such distribution is not explained in the current models. The estimated energy flux of the accelerated electrons corresponds to the threshold between the ``gentle'' and ``explosive'' evaporation regimes, according to the numerical simulations of \citet{Fisher85a,Fisher85b,Fisher85c}. However, the plasma evaporation velocities are substantially subsonic, meaning the evaporation process was of the ``gentle'' type.

At this point we do not give an explanation of this controversy. For better understanding of this process, detailed simulations of the evolution of profiles of the particular lines observed by IRIS are needed. Such type of work, for instance, was recently made by \citet{RubiodaCosta14}, who studied the radiation transfer in the H$\alpha$ 6563\,{\AA} and Ca\,II 8542\,{\AA} lines relative to observations of another M-class flare event.

\acknowledgments

The authors acknowledge the Big Bear Solar Observatory (BBSO) observing and technical team, and the IRIS mission team and the NASA Ames Research Center for their contributions and support. The work was partially supported by NASA grants NNX14AB68G, NNX14AB70G, and NNX11AO736; NSF grant AGS-1250818; and an NJIT grant.

\clearpage

\begin{deluxetable}{lc}
\tablecolumns{2}
\tablewidth{0pc}
\tablecaption{Fitting Parameters}
\tablehead{
\colhead{Model}   & \colhead{Fitting parameters}}
\startdata
\cutinhead{Time interval 21:03:20-21:05:24\,UT}
Thick-target    &   FL(E$>$E$_{low}$)$\sim{}10^{37}-10^{38}$\,electrons/s   \\
{}  &   E$_{low}<8$\,keV    \\
{}  &   $\delta{}=$7.6$\pm$0.1  \\
{}  &   E$_{high}=$28.5$\pm$0.6\,keV    \\
{}  &   $\chi^2=$4.28   \\
Thermal    &   EM$=$(1.9$\pm$0.1)$\times$10$^{46}$\,cm$^{-3}$   \\
{} &   T$=$24.9$\pm$1.1\,MK   \\
{}  &   $\chi^2=$2.71   \\
\cutinhead{Time interval 21:07:44-21:09:24\,UT}
Thermal-plus-nonthermal   &   FL(E$>$E$_{low}$)$=$(2.4$\pm$1.6)$\times{}10^{35}$\,electrons/s   \\
{}  &   E$_{low}=$13.0$\pm$1.3\,keV    \\
{}  &   $\delta{}=$8.1$\pm$0.6  \\
{}  &   E$_{high}=$31.0$\pm$3.9\,keV    \\
{}  &   EM$=$(5.4$\pm$4.6)$\times$10$^{49}$\,cm$^{-3}$   \\
{}  &   T$=$8.2$\pm$0.7\,MK   \\
{}  &   $\chi^2=$2.30   \\
Double-temperature thermal   &   EM$_{1}=$(3.52$\pm$1.39)$\times$10$^{49}$\,cm$^{-3}$  \\
{} &   T$_{1}=$8.45$\pm$0.34\,MK   \\
{} &   EM$_{2}=$(1.1$\pm$0.2)$\times$10$^{46}$\,cm$^{-3}$  \\
{} &   T$_{2}=$26.12$\pm$0.90\,MK   \\
{} &   $\chi^2=$2.34   \\
\cutinhead{Time interval 21:11:04-21:12:28\,UT}
Thermal-plus-nonthermal   &   FL(E$>$E$_{low}$)$=$(7.5$\pm$4.3)$\times{}10^{35}$\,electrons/s   \\
{}  &   E$_{low}=$12.8$\pm$1.1\,keV    \\
{}  &   $\delta{}=$7.3$\pm$0.4  \\
{}  &   E$_{high}=$26.9$\pm$0.7\,keV    \\
{}  &   EM$=$(12.5$\pm$10.1)$\times$10$^{49}$\,cm$^{-3}$   \\
{}  &   T$=$8.3$\pm$0.7\,MK   \\
{}  &   $\chi^2=$1.23   \\
Double-temperature thermal   &   EM$_{1}=$(13.7$\pm$4.9)$\times$10$^{49}$\,cm$^{-3}$  \\
{} &   T$_{1}=$8.1$\pm$0.2\,MK   \\
{} &   EM$_{2}=$(4.5$\pm$0.4)$\times$10$^{46}$\,cm$^{-3}$  \\
{} &   T$_{2}=$25.1$\pm$0.5\,MK   \\
{} &   $\chi^2=$1.96
\enddata
\end{deluxetable}

\begin{thebibliography}{}
\bibitem[Antiochos and Sturrock(1978)] {Antiochos78} {Antiochos}, S.~K. and {Sturrock}, P.~A. 1978, \apj, 220, 1137
\bibitem[Bornmann et al.(1996)] {Bornmann96} {Bornmann}, P.~L., {Speich}, D., {Hirman}, J. et al. 1996, Society of Photo-Optical Instrumentation Engineers (SPIE) Conference Series, 2812, 291
\bibitem[Brosius and Phillips(2004)] {Brosius04} {Brosius}, J.~W. and {Phillips}, K.~J.~H. 2004, \apj, 613, 580
\bibitem[Brosius and Holman(2007)] {Brosius07} {Brosius}, J.~W. and {Holman}, G.~D. 2007, \apjl, 659, L73
\bibitem[Brown(1971)] {Brown71} {Brown}, J.~C. 1971, \solphys, 18, 489
\bibitem[Brown(1973)] {Brown73} {Brown}, J.~C. 1973, \solphys, 31, 143
\bibitem[Chifor et al.(2008)] {Chifor08} {Chifor}, C., {Young}, P.~R., {Isobe}, H. et al. 2008, \aap, 481, L57
\bibitem[De Pontieu et al.(2014)] {DePontieu14} {De Pontieu}, B., {Title}, A.~M., {Lemen}, J.~R. et al. 2014, \solphys, 289,2733
\bibitem[Doshchek et al.(2013)] {Doschek13} {Doschek}, G.~A., {Warren}, H.~P. and {Young}, P.~R. 2013, \apj, 767, 55
\bibitem[Fisher et al.(1985a)] {Fisher85a} {Fisher}, G.~H., {Canfield}, R.~C. and {McClymont}, A.~N. 1985a, \apj, 289, 414
\bibitem[Fisher et al.(1985b)] {Fisher85b} {Fisher}, G.~H., {Canfield}, R.~C. and {McClymont}, A.~N. 1985b, \apj, 289, 425
\bibitem[Fisher et al.(1985c)] {Fisher85c} {Fisher}, G.~H., {Canfield}, R.~C. and {McClymont}, A.~N. 1985c, \apj, 289, 434
\bibitem[Fletcher et al.(2011)] {Fletcher11} {Fletcher}, L., {Dennis}, B.~R., {Hudson}, H.~S. et al. 2011, \ssr, 159, 19
\bibitem[Gudiksen et al.(2011)] {Gudiksen11} {Gudiksen}, B.~V., {Carlsson}, M., {Hansteen}, V.~H. et al. 2011, \aap, 531, 154
\bibitem[Kosovichev(1986)] {Kosovichev86} {Kosovichev}, A.~G. 1986, Bulletin Crimean Astrophysical Observatory, 75, 6
\bibitem[Kostiuk and Pekelner(1975)] {Kostiuk75} {Kostiuk}, N.~D. and {Pikelner}, S.~B. 1975, \sovast, 18, 590
\bibitem[Kurucz and Bell(1995)] {Kurucz95} {Kurucz}, R. and {Bell}, B. 1995, Atomic Line Data, Kurucz CD-ROM No.~23.~Cambridge, Mass.: Smithsonian Astrophysical Observatory
\bibitem[Leenaarts et al.(2012)] {Leenaarts12} {Leenaarts}, J., {Carlsson}, M. and {Rouppe van der Voort}, L. 2012, \apj, 749, 136
\bibitem[Leenaarts et al.(2013)] {Leenaarts13a} {Leenaarts}, J., {Pereira}, T.~M.~D., {Carlsson}, M., {Uitenbroek}, H., {De Pontieu}, B. 2013, \apj, 772, 90
\bibitem[Lemen et al.(2012)] {Lemen12} {Lemen}, J.~R., {Title}, A.~M., {Akin}, D.~J. et al. 2012, \solphys, 275, 17
\bibitem[Lin et al.(2002)] {Lin02} {Lin}, R.~P., {Dennis}, B.~R., {Hurford}, G.~J. et al. 2002, \solphys, 2010, 3
\bibitem[Liu et al.(2009)] {Liu09} {Liu}, W. and {Petrosian}, V. and {Mariska}, J.~T. 2009, \apj, 702, 1553
\bibitem[Livshits(1983)] {Livshits83} {Livshits}, M.~A. 1983, \sovast, 27, 557
\bibitem[Milligan et al.(2006)] {Milligan06} {Milligan}, R.~O., {Gallagher}, P.~T., {Mathioudakis}, M. and {Keenan}, F.~P., \apjl, 642, L169
\bibitem[Milligan and Dennis(2009)] {Milligan09} {Milligan}, R.~O. and {Dennis}, B.~R. 2009, \apj, 699, 968
\bibitem[Milligan(2015)] {Milligan15} {Milligan}, R.~O. 2015, ArXiv e-print 1501.04829
\bibitem[Raftery et al.(2009)] {Raftery09} {Raftery}, C.~L., {Gallagher}, P.~T., {Milligan}, R.~O. and {Klimchuk}, J.~A. 2009, \aap, 494, 1127
\bibitem[Rubio da Costa et al.(2014)] {RubiodaCosta14} {Rubio da Costa}, F., {Kleint}, L., {Petrosian}, V., {Sainz Dalda}, A. and {Liu}, W. 2014, ArXiv e-print 1412.1815
\bibitem[Scherrer et al.(2012)] {Scherrer12} {Scherrer}, P.~H., {Schou}, J., {Bush}, R.~I., {Kosovichev}, A.~G. et al. 2012, \solphys, 275, 207
\bibitem[Schmieder et al.(2014)] {Schmieder14} {Schmieder}, B., {Tian}, H., {Kucera}, T. et al. 2014, \aap, 569, A85
\bibitem[Schmelz et al.(2013)] {Schmelz13} {Schmelz}, J.~T., {Jenkins}, B.~S. and {Kimble}, J.~A. 2013, \solphys, 283, 325
\bibitem[Spitzer and H{\"a}rm(1953)] {Spitzer53} {Spitzer}, L. and {H{\"a}rm}, R. 1953, Physical Review, 89, 977
\bibitem[Sturrock(1968)] {Sturrock68} {Sturrock}, P.~A. 1968, International Astronomical Union Symposium 35, 471
\bibitem[Thomas et al.(1985)] {Thomas85} {Thomas}, R.~J., {Crannell}, C.~J. and {Starr}, R. 1985, \solphys, 95, 323
\bibitem[Tian et al.(2012)] {Tian12} {Tian}, H., {McIntosh}, S.~W., {Xia}, L., {He}, J., and {Wang}, X. 2012, \apj, 748, 106
\bibitem[Tian et al.(2014a)] {Tian14a} {Tian}, H., {DeLuca}, E., {Reeves}, K.~K. et al. 2014a, \apj, 786, 137
\bibitem[Tian et al.(2014b)] {Tian14b} {Tian}, H., {Li}, G., {Reeves}, K.~K. et al. 2014b, \apjl, 797, L14
\bibitem[Veronig and Brown(2004)] {Veronig04} {Veronig}, A.~M. and {Brown}, J.~C. 2004, \apjl, 603, L117
\bibitem[Vihlu et al.(2001)] {Vilhu01} {Vilhu}, O., {Muhli}, P., {Mewe}, R. and {Hakala}, P. 2001, \aap, 375, 492
\bibitem[Young et al.(2014)] {Young14} {Young}, P., {Tian}, H. and {Jaeggli}, S. 2014, ArXiv e-print 1409.8603
\bibitem[Zarro and Lemen(1988)] {Zarro88} {Zarro}, D.~M. and {Lemen}, J.~R. 1988, \apj, 329, 456
\end{thebibliography}
\end{document}